\documentclass[12pt]{article}
\usepackage[letterpaper,top=2cm,bottom=2cm,left=3cm,right=3cm,marginparwidth=1.75cm,margin=1in ]{geometry}
\usepackage{graphicx} 
\usepackage{graphbox}
\usepackage{float}
\usepackage{xcolor}
\usepackage{natbib}
\usepackage{amssymb}
\usepackage{hyperref}
\usepackage{multirow}
\usepackage{amsmath}
\usepackage{amsthm}
\newtheorem*{remark}{Remark}
\usepackage{comment}
\usepackage{url} 
\usepackage{soul}
\usepackage{ulem}
\usepackage{authblk}
\usepackage[shortlabels]{enumitem}
\DeclareMathOperator*{\argmin}{arg\,min}

\newcommand{\norm}[1]{\left\lVert#1\right\rVert}

\usepackage[ruled]{algorithm2e}
\usepackage{algpseudocode}

\title{Multivariate Planar Curves: A Statistical Framework for Shape Analysis in Images}

\author[1,2]{Issam-Ali Moindjié$^*$}
\author[1]{Cédric Beaulac}
\author[1]{Marie-Hélène Descary}
\affil[1]{Department of mathematics, Université du Québec à Montréal, Canada}
\affil[2]{LAMPS, Université de Perpignan Via Domitia, France}
\affil[*]{Corresponding author: issam-ali.moindjie@univ-perp.fr}

\date{ }

\begin{document}

\maketitle
\begin{abstract}

Recent developments in computer vision have made segmented images widely available across many domains, such as medicine, where segmented radiographs play an important role in diagnosis. As prediction problems are common in image analysis, this work explores the use of the object contours highlighted by such images as predictors in a supervised classification context. To this end, we develop a new statistical learning framework that accounts for the joint shape of the multiple objects contained in an image. We introduce a formalism that extends the study of a single random planar curve to the joint analysis of several planar curves, referred to as a multivariate planar curve. Modeling the contours jointly, rather than separately, preserves the inter-component information, such as their relative position, scale, and orientation, which is often essential to the analysis. Based on this model, we propose a joint alignment procedure and we extend core inferential tools to multivariate shapes: shape dissimilarity, Fréchet mean estimation, and tangent-space representation. These tangent coordinates are then used as predictors in standard functional classification models. A simulation study shows accurate recovery of deformation parameters over increasing noise levels. Then, through a cardiomegaly detection problem on segmented chest X-rays, we show that jointly modeling the contours is robust to misalignment and improves classification accuracy over both a contour-wise univariate analysis and a naive approach based on the raw curves.

\end{abstract}


\section{Introduction}\label{sec_intro}

Images constitute an important source of information in many scientific and industrial applications, including biomedical research \cite{lu2019} and road safety \cite{gu2020}, to name just a few. Due to acquisition technologies, images are naturally represented as matrices of pixels, where each pixel records information such as grayscale intensity or color values.

In many applications, the primary object of interest is not the collection of pixels itself, but rather the geometric structures represented in the image. In particular, the shape of objects often carries meaningful and interpretable information that may be difficult to extract from pixel-level representations alone. When the goal is to study geometry, a shape-based representation can provide a substantially more parsimonious description of the data while preserving the structural characteristics of the objects under investigation.

A common strategy for isolating objects and their shapes is image segmentation, whose goal is to partition an image into homogeneous regions sharing common characteristics such as texture, intensity, or gray level. Image segmentation has long been a central research area in pattern recognition and image analysis. Since the early developments of computer vision, numerous segmentation techniques have been proposed (see \cite{fu1981survey} for a survey). More recently, deep learning approaches have significantly improved segmentation performance in a wide range of applications (see, e.g., \cite{minaee2021}). These methods have demonstrated remarkable accuracy in practical settings, particularly in medicine, where segmentation plays an important role in image-based diagnosis and clinical decision-making \citep{wang2022}.

The output of a segmentation algorithm is typically a collection of masks identifying the objects of interest within an image. While such masks provide a natural pixel-based representation of shape, they remain discrete approximations of inherently continuous geometrical structures. Consequently, we argue that segmented images should not be analyzed solely through their masks, but rather through continuous representations of the shapes they contain. \cite{lelivre} advocate for a functional representation of shape that preserves the continuous nature of these objects, provides a parsimonious representation, and enables their analysis through the lens of Functional Data Analysis (FDA).

Building upon this idea, we propose to represent the multiple objects contained in an image as a multivariate planar curve. Each object is encoded by its contour, modeled as a planar closed curve described by two coordinate functions. An image containing multiple objects is therefore represented as a vector of planar curves, forming a multivariate functional observation. 

\begin{figure}[ht]
    \centering
    \begin{tabular}{c c c }
        (a) \includegraphics[height=4cm, align=c]{Figure_1/utils_example1.jpg}& &  (b) \begin{tabular}{c c c } 
              \includegraphics[scale=.08, align=c]{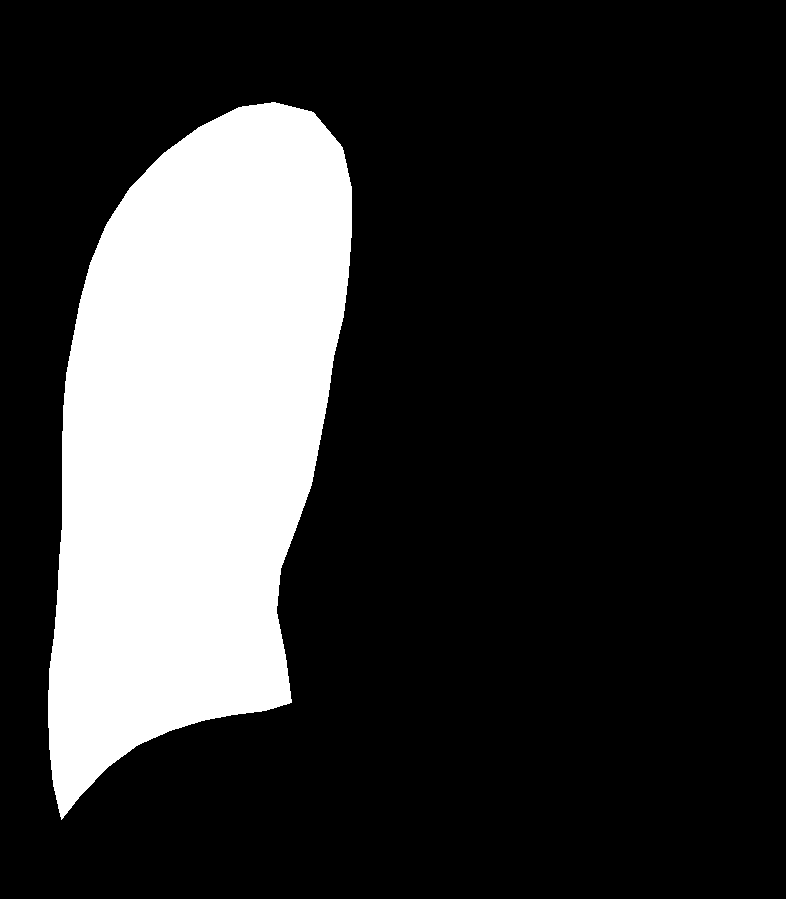} &
               \includegraphics[scale=.08, align=c]{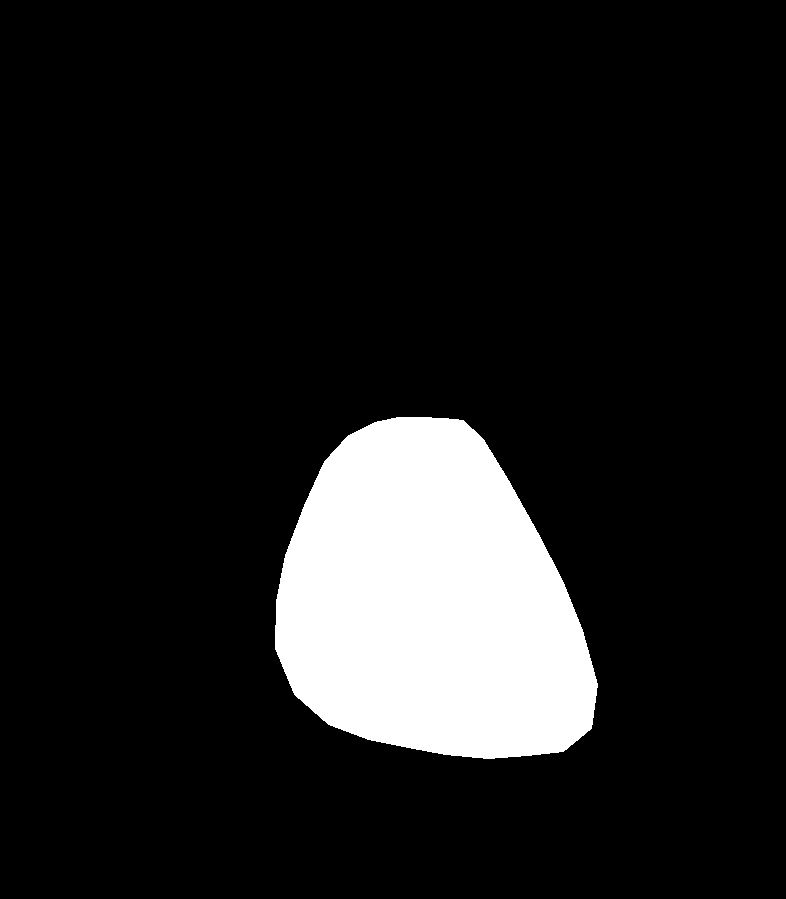} &
            \includegraphics[scale=.08, align=c]{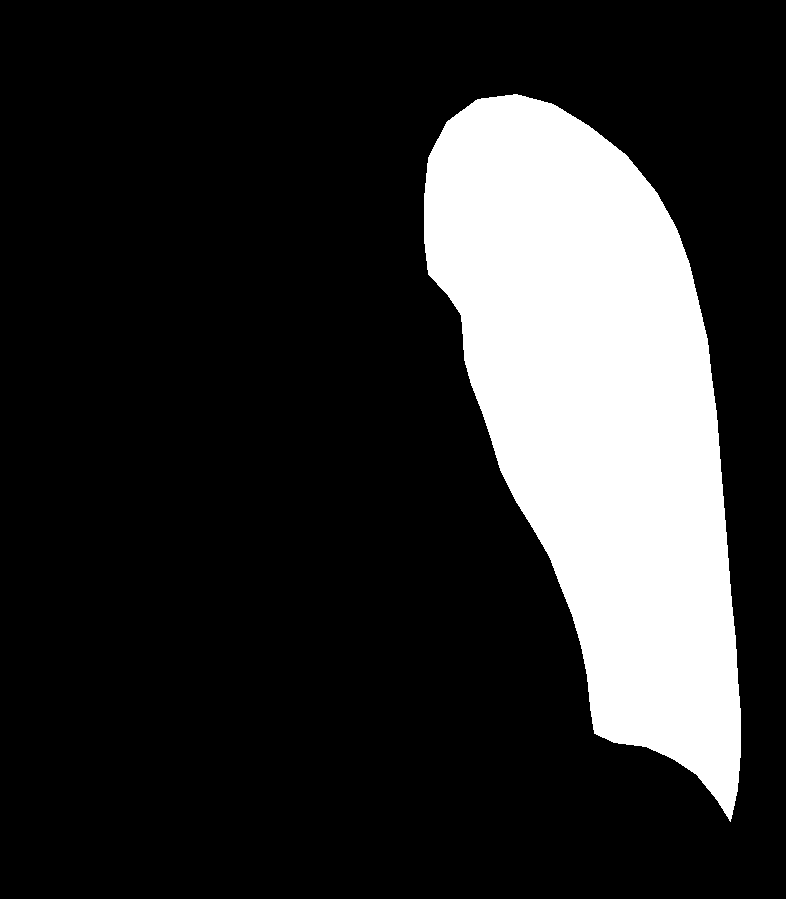} 
          \end{tabular} \\
              (c) \includegraphics[height=5cm, align=c]{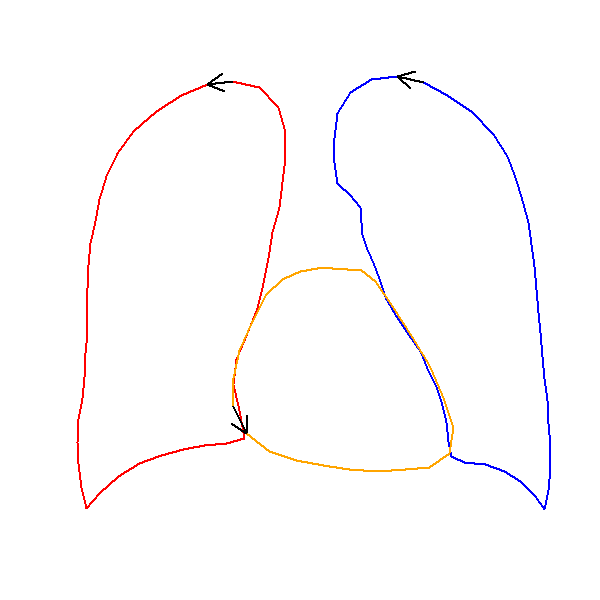} & & (d)\begin{tabular}{r c c}
              &$X$ & $Y$ \\ 
              $C_1$ &\includegraphics[align=c, scale=.1]{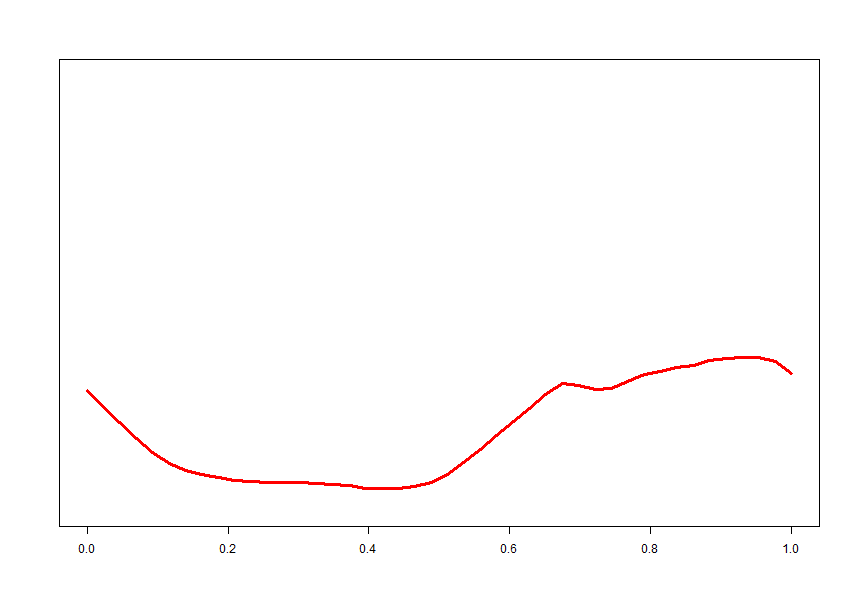} & \includegraphics[align=c, scale=.1]{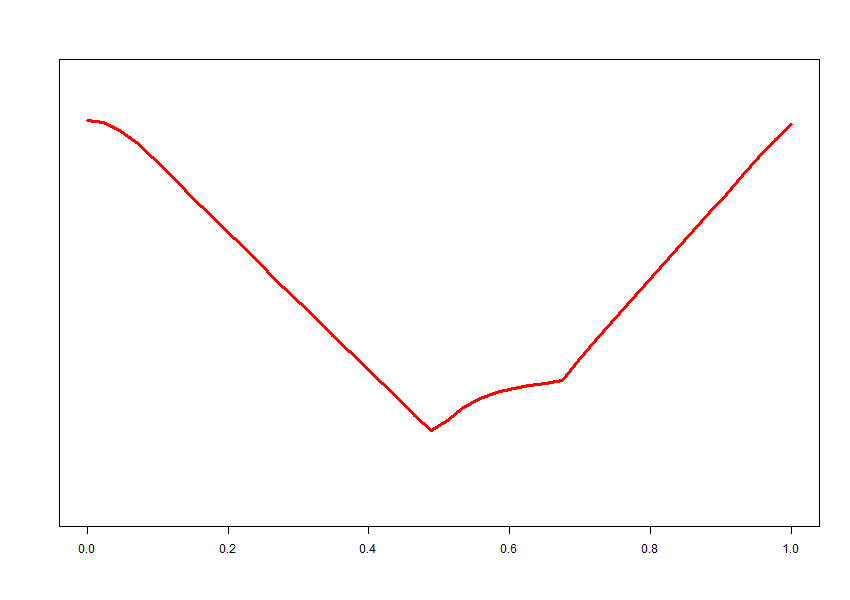} \\
              $C_2$ & \includegraphics[align=c, scale=.1]{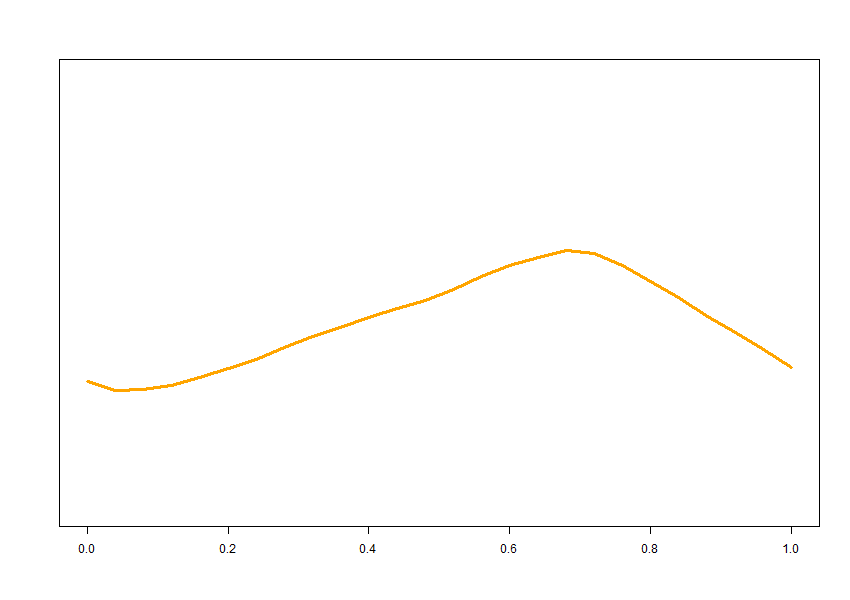} & \includegraphics[align=c, scale=.1]{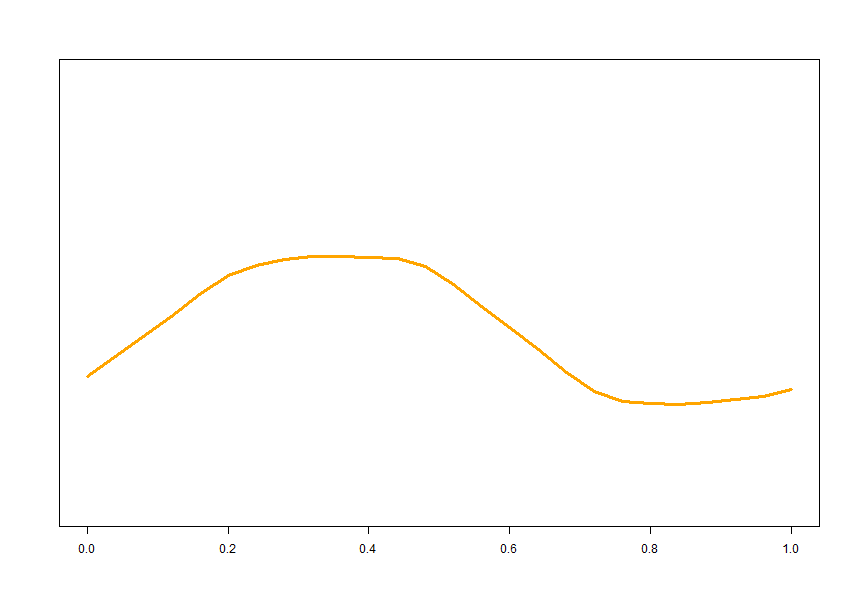} \\
              $C_3$ & \includegraphics[align=c, scale=.1]{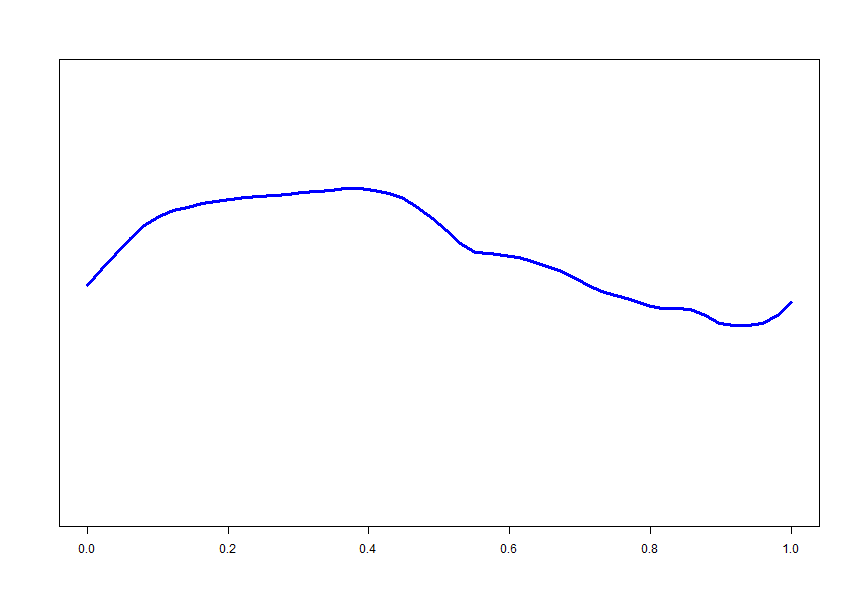} & \includegraphics[align=c, scale=.1]{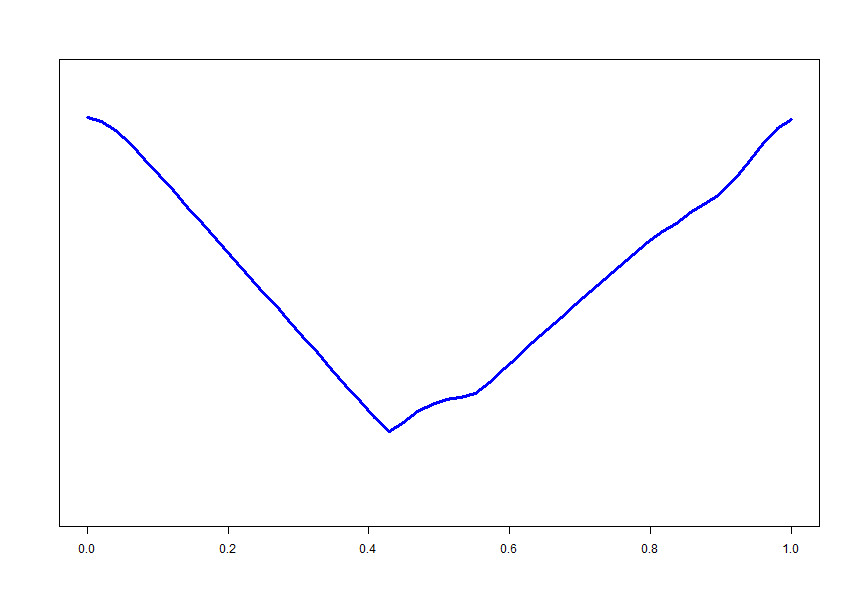} \\  \\       
         \end{tabular}  
      \\  
    \end{tabular}
    \caption{Representation of a segmented chest X-ray  \citep{gaggion2024chexmask} as a multivariate planar curve. 
(a) Original X-ray image. (b) Segmentation masks of the three organs of interest: right lung, heart, and left lung. (c) Contours that form the multivariate planar curve $\mathbf{C}$. (d) Coordinate functions $(X_j, Y_j)$ for each contour $C_j$ ($j=1,2,3$), obtained by traversing the contour in (c) from the starting point indicated by the black arrow.}
    \label{exs}
\end{figure}

As a motivating example, we consider segmented chest X-ray images containing the right lung, the heart, and the left lung, where the statistical problem is to detect cardiomegaly, a condition characterized by an abnormally large heart relative to the lungs. The relevant information is not contained in the shape of a single organ alone, but rather in the joint configuration of multiple anatomical structures. Figure \ref{exs} illustrates the proposed representation. Panel (a) shows the original chest X-ray, represented as a matrix of pixel intensities, while panel (b) displays the segmentation masks obtained using the method of \citet{gaggion2024chexmask}. Panels (c) and (d) present the objects that form the basis of our analysis. The contours of the right lung, heart, and left lung, denoted $C_1$, $C_2$, and $C_3$, are extracted from the masks and represented as planar closed curves. By traversing the boundary of each organ from the starting point indicated by the black arrow, each contour can be represented by two coordinate functions, namely $C_j^\top=(X_j,Y_j)$, $j=1,2,3$. Collectively, these contours form the multivariate planar curve
$\mathbf{C}=(C_1^\top,\ldots,C_p^\top)^\top$, with $p=3$, which constitutes the parsimonious image representation studied throughout this paper.

Statistical shape analysis has traditionally focused on objects represented by a single contour. In this setting, a shape is represented by a planar curve $C$, and a rich literature provides methodologies for registration, shape representation, similarity measurement, representative shape estimation, variability modeling, and classification (see, e.g., \cite{dryden1998}, \cite{srivastava_article}, \cite{younes1998}, \cite{lelivre}). The motivating example of Figure~\ref{exs}, however, departs from this single-contour setting, as the observation of interest is a collection of several contours rather than a single one. Analyzing such data calls for treating these contours jointly, as a single multivariate planar curve $\mathbf{C}=(C_1^\top,\ldots,C_p^\top)^\top$.

In this paper, we introduce a statistical framework for multivariate planar curves $\mathbf{C}$ with $p\geq 2$, extending classical shape analysis methodologies developed for single contours \citep{srivastava_article, dryden1998}. More precisely, we directly extend the work of \cite{axe1} to obtain a formal definition of multivariate shape together with associated deformation variables accounting for translation, rotation, scaling, and reparametrization effects.

Building upon this definition, we extend several fundamental tasks of statistical shape analysis identified by \cite{lelivre} to multivariate planar curves. In particular, we develop methodologies for: (i) alignment, (ii) quantifying similarities and dissimilarities, (iii) estimating an intrinsic mean and (iv) supervised classification.

In summary, the main contributions of this work are threefold:
\begin{itemize}
\item We introduce the notion of a multivariate planar curve and provide a formal definition of multivariate shape.

\item We develop a unified framework that extends several fundamental problems of statistical shape analysis from single contours to multivariate planar curves.

\item Through an application to the motivating dataset, we demonstrate the practical benefits of jointly modeling multiple interacting contours and show that this approach outperforms a naive extension of single-contour methodologies.

\end{itemize}

The rest of the paper is organized as follows. Section~\ref{sec_model_estim} recalls the univariate model of \cite{axe1}, introduces our multivariate model, and details the estimation of the shape variables, including task~(i), the joint alignment of the multiple contours. Section~\ref{sec_shape_analysis} then develops the statistical analysis of multivariate shapes: it equips the shape space with a dissimilarity measure between shapes (task~(ii)), summarizes a sample by its representative Fréchet mean (task~(iii)), and uses the resulting shape variables as predictors in shape-based classification (task~(iv)). Section~\ref{sec_num} evaluates the methodology through numerical experiments on synthetic data and an application to the cardiomegaly detection problem, and Section~\ref{sec_conclusion} concludes with a discussion of future works.

\section{Multivariate planar curves: model and estimation}\label{sec_model_estim}

In this section, we introduce our model for the multivariate random planar curve $\mathbf{C}$, together with the functional space it belongs to, and we describe the estimation of the deformation and shape variables. Since our construction builds upon the univariate case ($p=1$) of \cite{axe1}, we begin by briefly reviewing it.

\subsection{Univariate random planar closed curves} \label{sec_uniCurve}

In \cite{axe1}, a univariate random planar closed curve $C = (X, Y)^\top$ is defined as a bivariate function such that $C(t) = (X(t), Y(t))^\top$ for $t \in [0,1]$, with $C(0) = C(1)$. The functions $X$ and $Y$ represent the coordinates along the first and second axes, respectively, as the curve is traversed from a given starting location. Formally, $C$ is assumed to belong to the Hilbert space $\mathcal{H} = L_2([0, 1]) \times L_2([0, 1])$, equipped with the inner product
$$
\langle \boldsymbol{f}, \boldsymbol{g} \rangle_\mathcal{H} = \int_0^1 f_1(t) g_1(t) dt + \int_0^1 f_2(t) g_2(t) dt, \quad \text{for } \boldsymbol{f} = (f_1, f_2)^\top, \boldsymbol{g} = (g_1, g_2)^\top \in \mathcal{H},
$$
and we denote the induced norm by $\| \cdot \|_\mathcal{H}$.

The closed curve $C$ is modeled as a deformed version of a latent variable $\mathcal{C} \in \mathbf{S}^\infty$ according to
\begin{equation}
C(t) = \rho \ \mathbf{O}_\theta \ (\mathcal{C}\circ\gamma_\delta)(t) + \mathbf{T}, \quad t \in [0,1],
\label{eq_j}
\end{equation}
where $\mathbf{S}^\infty = \left\{ \boldsymbol{f} \in \mathcal{H} : \int_0^1 f_1(t) dt = \int_0^1 f_2(t) dt = 0, \ \| \boldsymbol{f} \|_\mathcal{H} = 1 \right\}$ denotes the centered unit Hilbert sphere. In this model, $\rho \in \mathbb{R}^+$ is a scaling factor, $\mathbf{O}_\theta \in SO(2)$ is a rotation matrix of angle $\theta \in [0, 2\pi]$, and $\mathbf{T} \in \mathbb{R}^2$ is a translation vector. 

In practice, the parametrization of a contour depends on the chosen starting point used to traverse the boundary. For instance, in Figure \ref{exs}(c), the black arrows indicate the starting points used to generate the coordinate functions shown in Figure \ref{exs}(d). Changing this starting point results in a different parametrization of the same geometric curve. In model \eqref{eq_j}, this effect is captured through the reparametrization function $\gamma_\delta$, which shifts the starting point along the contour. Indeed, we assume that $\gamma_\delta$ belongs to
$$
\Gamma=\left\{ \gamma_\delta : [0,1]\rightarrow [0,1], \ \gamma_\delta(t)= \text{mod}(t-\delta, 1), \ \delta \in [0,1] \right\},
$$
where $\text{mod}(\cdot,1)$ is the modulo 1 function. The function $\gamma_\delta$ can therefore be viewed as an extension of the classical time-shift warping function used in functional data registration to the case of planar closed curves, where the modulo operator accounts for the cyclic nature of closed curves. The latent variable $\mathcal{C}$ represents the shape of the curve $C$, that is, the geometric information that remains once translation, rotation and scaling effects have been removed \citep{dryden1998}.

\subsection{Multivariate random planar closed curves }

We now extend the univariate model of Section~\ref{sec_uniCurve} to the case where several contours are observed and analyzed jointly. Such an observation is modeled as a multivariate planar curve $\mathbf{C} = Vec(C_1, \ldots, C_p) = (C_1^\top, \ldots, C_p^\top)^\top$, where each component $C_j=(X_j,Y_j)^\top\in \mathcal{H}$ is a univariate planar closed curve. Thus, $\mathbf{C} =(X_1,Y_1,\ldots,X_p,Y_p)^\top\in \mathcal{H}^p$, where $\mathcal{H}^p$ denotes the product Hilbert space $\mathcal{H} \times \cdots \times \mathcal{H}$ ($p$ times), equipped with the inner product

$$
\langle \boldsymbol{f}, \boldsymbol{g} \rangle_{\mathcal{H}^p} = \sum_{j=1}^p \langle \boldsymbol{f}_j, \boldsymbol{g}_j \rangle_\mathcal{H}, \quad \text{for } \boldsymbol{f} = \begin{pmatrix}
\boldsymbol{f}_1 \\ 
\vdots \\ 
\boldsymbol{f}_p
\end{pmatrix}, \ \boldsymbol{g} = \begin{pmatrix}
\boldsymbol{g}_1 \\ 
\vdots \\ 
\boldsymbol{g}_p
\end{pmatrix} \in \mathcal{H}^p,
$$
with $\boldsymbol{f}_j=(f_{j1},f_{j2})^\top$ and $\boldsymbol{g}_j=(g_{j1},g_{j2})^\top$. We denote by $\| \cdot \|_{\mathcal{H}^p}$ the norm induced by this inner product.

In order to establish a reasonable model, the key modeling question is to determine which geometric variations should be regarded as deformations caused by the image acquisition process and which should instead be considered meaningful characteristics of the multivariate shape. Since all contours originate from the same image, translation, rotation, and scaling should be viewed as global deformations affecting the image as a whole rather than each contour individually. Conversely, the relative position, orientation, and size of the objects are intrinsic characteristics of the multivariate shape itself and should therefore be encoded in the latent shape variable.

Motivated by this observation, we propose the following model:
\begin{equation}
C_j = \rho \, \mathbf{O}_\theta \, \tilde{C}_j \circ \gamma_{\delta_j} + \mathbf{T}, \quad j = 1, \ldots, p,
\label{eq_mvc}
\end{equation}
where the scaling factor $\rho$, rotation angle $\theta$, and translation vector $\mathbf{T}$ are global deformations shared across all contours, while each component retains its own reparametrization function $\gamma_{\delta_j}$. Allowing contour-specific reparametrizations is natural since the starting point used to parameterize each contour depends on the contour extraction procedure and is independent across objects. In contrast, the remaining deformation parameters arise from the acquisition process and therefore affect every contour simultaneously.

Under model \eqref{eq_mvc}, the multivariate shape is not defined as a collection of independent shapes, but rather as the geometric structure formed by all contours after removing only the deformations that act globally on the image. As a result, relative variations between contours, such as the size, orientation, and position of one object with respect to another, are encoded directly in the multivariate shape variable $\tilde{\mathbf{C}} = Vec(\tilde{C}_1, \ldots, \tilde{C}_p)$.

For comparison, a straightforward extension of the univariate model \eqref{eq_j} would consist in modeling each contour independently by defining
\begin{equation}\label{eq_unie}
C_j = \rho_j \, \mathbf{O}_{\theta_j} \, \mathcal{C}_j \circ \gamma_{\delta_j} + \mathbf{T}_j, \quad j = 1, \ldots, p,
\end{equation}
where each contour has its own deformation parameters. Although this formulation is a natural extension of the univariate setting, it implicitly removes the relative position, orientation, and scale of the objects from the multivariate shape by capturing them into contour-specific deformation parameters.  Returning to the motivating example of Section~\ref{sec_intro}, such a model would treat the heart and lungs independently. Consequently, the relative size of the heart with respect to the lungs would be treated as a deformation rather than a statistically meaningful characteristic of the multivariate shape. As a result, some important geometric patterns would no longer be encoded in the latent shape variable.

The global shape of $\mathbf{C}$ is therefore represented by the latent multivariate planar curve $\tilde{\mathbf{C}} = Vec(\tilde{C}_1, \ldots, \tilde{C}_p) \in \mathbf{S}^\infty_p $, as defined in \eqref{eq_mvc}, where $\mathbf{S}^\infty_p$ denotes the centered unit Hilbert sphere in $\mathcal{H}^p$:
$$
\mathbf{S}^\infty_p = \left\{ \boldsymbol{f} \in \mathcal{H}^p : \sum_{j=1}^p \int_0^1 f_{j1}(t) dt = \sum_{j=1}^p \int_0^1 f_{j2}(t) dt = 0, \ \| \boldsymbol{f} \|_{\mathcal{H}^p} = 1 \right\}.
$$
Model \eqref{eq_mvc} can be written in the following compact vectorized form:
\begin{equation}
\mathbf{C} = \rho (\mathbf{I}_p \otimes\mathbf{O}_\theta)  \, \tilde{\mathbf{C}} \boldsymbol{\circ} \boldsymbol{\gamma} + \left( \mathbf{1}_p \otimes \mathbf{T} \right),
\label{eq_gen}
\end{equation}
where $\otimes$ denotes the Kronecker product, $\mathbf{I}_p$ is the identity matrix of dimension $p\times p$, $\mathbf{1}_p$ is the $p$-dimensional vector of ones, and $\boldsymbol{\gamma} = (\gamma_{\delta_1}, \ldots, \gamma_{\delta_p}) \in \Gamma^p$ is the vector of reparametrization functions, the composition being defined component-wise as $\tilde{\mathbf{C}} \boldsymbol{\circ} \boldsymbol{\gamma} = (\tilde{C}_1 \circ \gamma_{\delta_1}, \ldots, \tilde{C}_p \circ \gamma_{\delta_p})^\top$. As in the univariate setting, $(\Gamma^p, \circ)$ is a commutative group and isometry property
$$
\| \boldsymbol{f} \|_{\mathcal{H}^p} = \left\| \left( \mathbf{I}_p \otimes \mathbf{O}_\theta \right) \, \boldsymbol{f} \circ \boldsymbol{\gamma} \right\|_{\mathcal{H}^p}, \textrm{ for } \boldsymbol{f} \in \mathcal{H}^p \textrm{ and } \boldsymbol{\gamma} \in \Gamma^p,
$$
holds; this result follows directly from Proposition 2.1 in \cite{axe1}.

\subsection{Functional estimation}\label{sec_functional_estim}\label{sec_estim_shape}

We now describe how to recover the shape $\tilde{\mathbf{C}}$ from a curve $\mathbf{C}$. Inverting the deformation model \eqref{eq_mvc}, each component of the shape can be written as
\begin{equation} \label{shape_j}
\tilde C_j=\mathbf{O}_{\theta}^\top\left(\tfrac{1}{\rho}(C_j-\mathbf{T})\right) \circ \gamma_{1-\delta_j}=\mathbf{O}_{\theta}^\top\left(C_j^*\right) \circ \gamma_{1-\delta_j}, \quad j=1,\ldots,p,
\end{equation}
where $\gamma_{1-\delta_j}$ is the inverse of $\gamma_{\delta_j}$ \citep{axe1} and $\mathbf{C}^*=Vec(C_1^*,\ldots,C_p^*) \in \mathbf{S}^\infty_p$ is the \textit{pre-shape} of $\mathbf{C}$, that is, the curve once translation and scaling have been removed. This inversion is purely algebraic, so recovering the shape amounts to estimating the deformation parameters $\mathbf{T}$, $\rho$, $\theta$ and $\boldsymbol{\delta}=(\delta_1,\ldots,\delta_p)$.

To carry out this estimation in practice, we follow the standard functional data analysis approach and represent the curve in a finite Fourier basis. Each component admits the representation
\begin{equation}
    C_j(t)= \begin{pmatrix} X_j(t)\\  Y_j(t) \end{pmatrix} =\begin{pmatrix} B_{j1}\\  B_{j2} \end{pmatrix}+ \sum_{m=1}^M  \begin{pmatrix} A_{jm1}\\  A_{jm2} \end{pmatrix} \phi_m(t) = {B}_j + {A}_j \boldsymbol{\phi}(t), \quad j=1,\ldots, p,
    \label{eqFourrier1}
\end{equation}
where ${B}_j\in \mathbb{R}^{2}$, ${A}_j\in \mathbb{R}^{2\times M}$ are coefficient matrices and $\boldsymbol{\phi}(t)=( \phi_1(t) , \ldots, \phi_M(t))^\top$ contains the first $M$ Fourier basis functions ($M$ even)
\begin{equation*}\label{fourier}
\phi_{k}(t)=\left\{ \begin{array}{ll} \sqrt{2}\sin((k+1)\pi t) & \textrm{if $k$ is odd,}   \\ \sqrt{2}\cos(k\pi t) & \textrm{if $k$ is even.} \end{array} \right.
\end{equation*}
In practice, a curve $\mathbf{C}$ is not observed as a continuous function but only on a finite grid of points, typically dictated by the image resolution. The coefficient matrices ${B}_j$ and ${A}_j$ are therefore unknown and are estimated from these observations by smoothing, for instance by minimizing a least-squares criterion between the expansion \eqref{eqFourrier1} and the observed values. Stacking these coefficients, we write the multivariate curve compactly as
\begin{equation}
  \mathbf{C}=  \mathbf{B}+\mathbf{A}  \boldsymbol{\phi},
    \label{eq1}
\end{equation}
where $\mathbf{B}=Vec({B}_1,\ldots,{B}_p) \in \mathbb{R}^{2p}$ and $\mathbf{A} \in \mathbb{R}^{2p\times M}$ is the matrix formed by vertically concatenating ${A}_1, \ldots, {A}_p$. Building on this representation, Section~\ref{estim_T_rho} computes the translation and scaling parameters (hence the pre-shape $\mathbf{C}^*$) and Section~\ref{ficp} estimates the rotation and reparametrization that map $\mathbf{C}^*$ to the shape $\tilde{\mathbf{C}}$.

\subsubsection{Translation and scaling parameters}\label{estim_T_rho}

Since $\tilde{\mathbf{C}}$ is centered and has unit norm, the translation and scaling factors are identified directly from model \eqref{eq_gen}, in closed form, as
\begin{equation}
\label{def_T_rho}
\mathbf{T} =\frac{1}{p} \sum_{j=1}^p \int_0^1 C_j(t)\,dt = \left(\frac{1}{p} \sum_{j=1}^p \int_0^1 X_j(t)\,dt,\ \frac{1}{p} \sum_{j=1}^p \int_0^1 Y_j(t)\,dt\right)^\top, \ \rho = \norm{ \mathbf{C}- \mathbf{1}_p \otimes \mathbf{T}}_{\mathcal{H}^p}.
\end{equation}
The translation $\mathbf{T}$ is thus the centroid of the multivariate curve (its average position in the image plane) while the scaling $\rho$ is its overall size, measured as the distance from $\mathbf{C}$ to that centroid. 

In practice, these two parameters are evaluated through the Fourier representation \eqref{eq1}. Since the basis functions are orthonormal and satisfy $\int_0^1 \phi_m(t)\,dt=0$, the integral and the norm in \eqref{def_T_rho} reduce to the coefficient expressions
\begin{equation*}
\mathbf{T}= \frac{1}{p}\sum_{j=1}^p {B}_j \qquad \textrm{and} \qquad  \rho = \sqrt{ \norm{\mathbf{A}}_F^2+\norm{\mathbf{B}-\mathbf{1}_p \otimes \mathbf{T}}_F^2 },
\end{equation*}
where $\norm{\cdot}_F$ is the Frobenius norm. Removing these two effects from $\mathbf{C}$ yields the pre-shape $\mathbf{C}^*$ of \eqref{shape_j}, which in the Fourier basis reads
$$
\mathbf{C}^*= \mathbf{B}^* +  \mathbf{A}^* \boldsymbol{\phi}, \quad \textrm{with} \quad \mathbf{A}^*=  \tfrac{1}{\rho}\mathbf{A}  \quad \textrm{and} \quad \mathbf{B}^*= \tfrac{1}{\rho}\left(\mathbf{B}- \mathbf{1}_p \otimes \mathbf{T}\right).
$$
The pre-shape $\mathbf{C}^*$ lies in $\mathbf{S}^\infty_p$ but still carries the rotation and reparametrization effects, which are estimated in the next section.

\subsubsection{Multivariate contour alignment}\label{ficp}

Recovering the shape from the pre-shape is an alignment problem: we estimate the rotation and reparametrization that best align a pre-shape $\mathbf{C}^*$ to a template $\bar{\mathbf{C}}=Vec(\bar C_1,\ldots,\bar C_p)\in\mathbf{S}_p^\infty$. The choice of this template is discussed in Section~\ref{sec_shape_analysis}; here it is taken as given. Aligning $\mathbf{C}^*$ to $\bar{\mathbf{C}}$ amounts to estimating the rotation parameter $\theta$ and the reparametrization parameter $\boldsymbol{\delta}$ that solve
\begin{equation}
    \min_{\boldsymbol{\delta}\in [0, 1]^p,\ \theta \in [0, 2\pi] }\sum_{j=1}^p\norm{\mathbf{O}_\theta {\bar{C}}_j\circ{\gamma}_{\delta_j}-{C}^*_j}_\mathcal{H}^2.
    \label{pbl}
\end{equation}
For a single contour ($p=1$), \cite{axe1} solve \eqref{pbl} with the \textit{Iterative Closest Function} (ICF) algorithm, an alternating optimization procedure, inspired by the iterative closest point (ICP) algorithm \citep{ICP-review}, that operates directly in the functional space without discretizing the curves. Since the dimension of the search space in \eqref{pbl} grows with the number of contours $p$, optimizing jointly over $(\theta,\delta_1,\ldots,\delta_p)$ is impractical. We therefore extend the ICF algorithm to the multivariate setting: starting from an initial value of $\boldsymbol{\delta}$, it alternates the two steps below until convergence. Both are the direct multivariate counterparts of the steps of \cite{axe1}; the only structural difference is that the shared rotation $\theta$ is now estimated jointly from all $p$ components.

\begin{enumerate}
\item[(i)] \textbf{Estimation of $\theta$ for a fixed $\boldsymbol{\delta}$.} The shared rotation angle solves the Procrustes problem
$$
\hat{\theta}=\argmin_{\theta\in [0, 2\pi]}\sum_{j=1}^p\norm{\mathbf{O}_\theta \bar{C}_j\circ \gamma_{\delta_j}-C_j^*}_\mathcal{H}^2 .
$$
As in the univariate case \citep{axe1}, the minimizer belongs to $\{\theta_1,\theta_1+\pi\}$, where
$$
\tan(\theta_1)=\frac{ \sum_{j=1}^p \left\{ \langle Y_j^*,\bar{X}_j\circ \gamma_{\delta_j} \rangle_{L_2} - \langle X_j^*,\bar{Y}_j\circ \gamma_{\delta_j} \rangle_{L_2}\right\}}{\sum_{j=1}^p \left\{ \langle X_j^*,\bar{X}_j\circ \gamma_{\delta_j} \rangle_{L_2} + \langle Y_j^*,\bar{Y}_j\circ \gamma_{\delta_j} \rangle_{L_2}\right\}},
$$
with $\bar{C}_j=(\bar{X}_j, \bar{Y}_j)^\top$, $C_j^*=(X_j^*, Y_j^*)^\top$, and $\langle \cdot, \cdot\rangle_{L_2}$ the inner product of $L_2([0,1])$. The two candidates correspond to opposite orientations; we evaluate the objective at both and keep the one with the smaller value.

\item[(ii)] \textbf{Estimation of $\boldsymbol{\delta}$ for a fixed $\theta$.} For a fixed rotation, the objective in \eqref{pbl} is a sum of terms each depending on a single $\delta_j$. The reparametrization parameters therefore decouple across components, and each is obtained by solving
\begin{equation}
\hat{\delta}_j=\argmin_{\delta_j\in [0,1]}\norm{\mathbf{O}_\theta \bar{C}_j\circ \gamma_{\delta_j}-C_j^*}_\mathcal{H}^2 , \quad j=1,\ldots,p.
\label{pb2}
\end{equation}
Each subproblem \eqref{pb2} is exactly the univariate alignment problem of \cite{axe1}; as there, it has no closed-form solution and each $\hat{\delta}_j$ is computed numerically by a bisection method.
\end{enumerate}

\begin{remark}
As in the univariate case, the multivariate ICF algorithm is sensitive to the initialization of $\boldsymbol{\delta}$. In practice, we run it from several initial values and retain the pair $(\hat{\theta},\hat{\boldsymbol{\delta}})$ that minimizes the objective in \eqref{pbl}.
\end{remark}
Once $(\hat{\theta},\hat{\boldsymbol{\delta}})$ have been obtained, the shape $\tilde{ \mathbf{C}}$ is recovered through the inverse transformations \eqref{shape_j}.

\section{Statistical analysis of multivariate shapes}\label{sec_shape_analysis}

This section develops the statistical analysis of multivariate shapes and their use as predictors in a classification task. Throughout, we consider a sample of $n$ multivariate planar curves $\mathbf{C}_1,\ldots,\mathbf{C}_n$, for instance the $n$ segmented images of a dataset, from which the pre-shapes $\mathbf{C}_1^*,\ldots,\mathbf{C}_n^*$ are obtained as described in Section~\ref{sec_functional_estim}.

By construction, each pre-shape $\mathbf{C}_i^*$ lies on the sphere $\mathbf{S}^\infty_p$, but still differs from the shape $\tilde{\mathbf{C}}_i$ of $\mathbf{C}_i$ by a rotation and a reparametrization. The shape $\tilde{\mathbf{C}}_i$ is what remains once these last two deformations are removed as well; equivalently, it is identified with the equivalence class of $\mathbf{C}_i^*$ under the relation
\begin{equation*}
\boldsymbol{f} \sim \boldsymbol{g} \iff \exists\, (\theta, \boldsymbol{\gamma}) \in [0, 2\pi] \times \Gamma^p \text{ such that } \boldsymbol{f} = (\mathbf{I}_p \otimes \mathbf{O}_\theta)\, \boldsymbol{g} \boldsymbol{\circ} \boldsymbol{\gamma}, \qquad \boldsymbol{f},\boldsymbol{g}\in\mathbf{S}^\infty_p,
\end{equation*}
which holds when $\boldsymbol{f}$ and $\boldsymbol{g}$ coincide up to a rotation and a reparametrization. The shapes thus belong to the quotient \textit{shape space} $\mathbf{S}^\infty_p/\!\sim$, while $\mathbf{S}^\infty_p$ itself is called the \textit{pre-shape space}, as it contains each shape together with all its rotated and reparametrized versions. This shape space is nonlinear, so classical statistical tools, which rely on a linear structure, cannot be applied to the shapes directly.

We address this difficulty through the standard strategy of statistical shape analysis (see e.g.\ \cite{dryden1998}, \cite{srivastava}), which proceeds in three steps. First, we summarize the sample by an intrinsic mean shape, the Fréchet mean $\boldsymbol{\mu}$ (Section~\ref{functional_mean}). Second, we linearize the shape space in the neighborhood of $\boldsymbol{\mu}$ by projecting the shapes onto the tangent space at $\boldsymbol{\mu}$ (Section~\ref{sec_tangent}). Finally, we perform the statistical analysis, here the shape-based classification, in this linear space (Section~\ref{sec_classi}).

\subsection{The Fréchet mean of multivariate shapes}\label{functional_mean}

To compare two shapes $\tilde{\mathbf{C}}_1$ and $\tilde{\mathbf{C}}_2$, that is, the equivalence classes of their pre-shapes $\mathbf{C}_1^*$ and $\mathbf{C}_2^*$ under the relation introduced above, the distance $d_{\mathcal{H}^p}$ (the distance induced by $\norm{\cdot}_{\mathcal{H}^p}$) is minimized over the rotation and reparametrization that align one pre-shape to the other:
\begin{equation}
d_{\mathcal{S}}(\tilde{\mathbf{C}}_1, \tilde{\mathbf{C}}_2) = \min_{\theta\in [0, 2\pi],\ \boldsymbol{\gamma}\in \Gamma^p} d_{\mathcal{H}^p}\!\left( \mathbf{C}_1^*,\ (\mathbf{I}_p \otimes \mathbf{O}_\theta)\, \mathbf{C}_2^* \boldsymbol{\circ} \boldsymbol{\gamma} \right).
\label{shape_dist}
\end{equation}
The minimization in \eqref{shape_dist} is the alignment problem \eqref{pbl} applied to two pre-shapes (the distance minimized in \eqref{pbl} and the distance $d_{\mathcal{H}^p} $ in \eqref{shape_dist} share the same minimizer on $\mathbf{S}^\infty_p$), so the multivariate ICF algorithm of Section~\ref{ficp} computes $d_{\mathcal{S}}$. By construction, $d_{\mathcal{S}}$ is a dissimilarity measure between multivariate shapes that is invariant to translation, scaling, rotation and reparametrization: it is nonnegative and symmetric, and vanishes if and only if the two curves share the same shape.

Following the classical approach in statistical shape analysis \citep{dryden1998}, we summarize the shape variable $\tilde{\mathbf{C}}$ by its Fréchet mean
\begin{equation}
\boldsymbol{\mu} = \argmin_{\boldsymbol{f}\in \mathbf{S}^\infty_p} \mathbb{E}\!\left[\, d_{\mathcal{H}^p}^2(\tilde{\mathbf{C}}, \boldsymbol{f})\,\right] \in\ \mathbf{S}^\infty_p,
\label{frechet_mean}
\end{equation}
which is the shape that is, on average, closest to the random shape $\tilde{\mathbf{C}}$, and thus serves as a natural representative of the population. It also settles the template choice left open in Section~\ref{ficp}: as is standard in statistical shape analysis \citep{dryden1998}, each pre-shape is aligned to the mean shape itself, which places the whole sample in a common frame.

The Fréchet mean \eqref{frechet_mean} has no closed-form expression and must be estimated from the sample. This estimation is itself circular, since the template above is the very mean we seek: $\boldsymbol{\mu}$ is needed to align the shapes, while the aligned shapes are needed to compute $\boldsymbol{\mu}$. Following \cite{axe1}, we break this dependency with an iterative procedure\label{sec_iter_algo} that, starting from an initial template, repeats the two steps
\begin{enumerate}[noitemsep]
\item[(a)] \textit{alignment:} for $i=1,\ldots,n$, align the pre-shape $\mathbf{C}_i^*$ to the current template with the ICF algorithm of Section~\ref{ficp}, producing the shapes $\hat{\tilde{\mathbf{C}}}_1,\ldots,\hat{\tilde{\mathbf{C}}}_n$;
\item[(b)] \textit{update:} replace the template by the empirical Fréchet mean of $\hat{\tilde{\mathbf{C}}}_1,\ldots,\hat{\tilde{\mathbf{C}}}_n$, that is, the sample counterpart of \eqref{frechet_mean},
\end{enumerate}
until convergence. The procedure then returns the estimated Fréchet mean $\hat{\boldsymbol{\mu}}$ together with the aligned shapes $\hat{\tilde{\mathbf{C}}}_1,\ldots,\hat{\tilde{\mathbf{C}}}_n$. Both the empirical Fréchet mean of step~(b) and the overall procedure are the multivariate counterparts of those in \cite{axe1}, to which we refer for the numerical details.

\subsection{The tangent space}\label{sec_tangent}

The classical tangent projection (\cite{dryden1998}, \cite{lelivre}, \cite{dai2022}) can be extended in our case. We recall that the tangent space to $\mathbf{S}^\infty_p$ at $\boldsymbol{\mu}$ is the linear subspace
\begin{equation*}
\mathcal{T}_{\boldsymbol{\mu}} = \left\{ \boldsymbol{f}\in \mathcal{H}^p : \langle \boldsymbol{f}, \boldsymbol{\mu} \rangle_{\mathcal{H}^p} = 0 \right\}.
\end{equation*}
 Hence, equipped with the Fréchet mean, we can linearize the shape space in its neighborhood. Indeed, unlike $\mathbf{S}^\infty_p$, $\mathcal{T}_{\boldsymbol{\mu} }$ is closed under linear combinations and therefore amenable to classical (functional) linear methods. Shapes are mapped to $\mathcal{T}_{\boldsymbol{\mu}}$, and back, through the logarithm and exponential maps, whose well-established properties make them a standard choice in this context (see e.g.\ \cite{dryden1998}, \cite{dai2022}). The logarithm map $\mathcal{L}_{\boldsymbol{\mu}}: \mathbf{S}^\infty_p\setminus\{-\boldsymbol{\mu}\} \to \mathcal{T}_{\boldsymbol{\mu}}$ projects a shape onto the tangent space,
\begin{equation}
\mathcal{L}_{\boldsymbol{\mu}}(\boldsymbol{f}) = \frac{\omega}{\sin(\omega)}\left( \boldsymbol{f} - \cos(\omega)\, \boldsymbol{\mu} \right), \qquad \omega = d_{\mathbf{S}^\infty_p}(\boldsymbol{f}, \boldsymbol{\mu}),
\label{log_map}
\end{equation}
whereas the exponential map $\mathrm{Exp}_{\boldsymbol{\mu}}: \mathcal{T}_{\boldsymbol{\mu}} \to \mathbf{S}^\infty_p$ maps a tangent vector back onto the sphere,
\begin{equation}
\mathrm{Exp}_{\boldsymbol{\mu}}(\boldsymbol{v}) = \cos\!\left(\norm{\boldsymbol{v}}_{\mathcal{H}^p}\right) \boldsymbol{\mu} + \sin\!\left(\norm{\boldsymbol{v}}_{\mathcal{H}^p}\right) \frac{\boldsymbol{v}}{\norm{\boldsymbol{v}}_{\mathcal{H}^p}}, \qquad \boldsymbol{v}\in\mathcal{T}_{\boldsymbol{\mu}}.
\label{exp_map}
\end{equation}
The two maps are mutual inverses in a neighborhood of $\boldsymbol{\mu}$, so that $\mathrm{Exp}_{\boldsymbol{\mu}}\big(\mathcal{L}_{\boldsymbol{\mu}}(\boldsymbol{f})\big) = \boldsymbol{f}$: one can therefore carry out the statistical analysis on the tangent coordinates $\mathcal{L}_{\boldsymbol{\mu}}(\boldsymbol{f})$ and, when needed, map the results back onto the sphere through $\mathrm{Exp}_{\boldsymbol{\mu}}$.

In practice, each sample shape is represented by its tangent coordinate $\mathcal{L}_{\boldsymbol{\mu}}(\hat{\tilde{\mathbf{C}}}_i)$, $i=1,\ldots,n$. This linearization is accurate as long as the shapes lie close to the reference point: choosing $\boldsymbol{\mu}$ as the Fréchet mean, which minimizes the average squared distance \eqref{frechet_mean}, keeps the shapes as close as possible to $\boldsymbol{\mu}$ and thus makes the projection as faithful as possible.

\subsection{Classification of multivariate shapes}\label{sec_classi}

Because the tangent space $\mathcal{T}_{\boldsymbol{\mu}}$ is linear, the projected shapes can serve as predictors in standard functional classification models, as is common in statistical shape analysis \citep[Chap.~13]{dryden1998}. Formally, let $Y$ be a binary response associated with a multivariate planar curve $\mathbf{C}$. Using the tangent coordinate $\mathcal{L}_{\boldsymbol{\mu}}(\tilde{\mathbf{C}})$ as a predictor casts the problem into the well-studied framework of functional classification with multivariate functional predictors (see e.g.\ \cite{godwin2013}, \cite{mfpls}, \cite{pcr}). A typical instance is the functional linear model
\begin{equation}\label{model_classi}
Y=\psi\!\left(\beta_0+ \langle \mathcal{L}_{\boldsymbol{\mu}}(\tilde{\mathbf{C}}), \boldsymbol{\beta} \rangle_{\mathcal{H}^p}\right),
\end{equation}
where $\beta_0 \in \mathbb{R}$ and $\boldsymbol{\beta}\in \mathcal{H}^p$ are estimated from the data and $\psi: \mathbb{R} \to \{0, 1\}$ is a link function determined by the classification method, such as linear discriminant analysis or logistic regression. The specific methods used in our experiments are presented in Section~\ref{class}.

\section{Numerical experiments}\label{sec_num}

In this section, we present numerical experiments designed to assess the performance of the proposed methodology. Our primary objective is to evaluate its effectiveness in a classification setting. To this end, we apply it to the real-world task of detecting cardiomegaly from chest X-ray images. Before considering this application, we first study the accuracy of the proposed alignment procedure on a synthetic dataset. This preliminary analysis is important because classification of shapes strongly depends on the accurate estimation of the deformation parameters, in particular the rotation and reparametrization components.

\subsection{Performance study of the alignment procedure}
\label{simu1}

We conduct a simulation study to numerically assess the accuracy of the deformation estimation described in Section~\ref{sec_estim_shape}. Since translation and scaling parameters are known to be accurately estimated in practice \citep{axe1}, this performance study focuses exclusively on the estimation of the rotation and reparametrization parameters. In other words, we evaluate the ability of our proposed alignment procedure to recover the underlying shapes from their corresponding pre-shapes.

\begin{figure}[ht]
    \centering
    \includegraphics[width=0.35\linewidth]{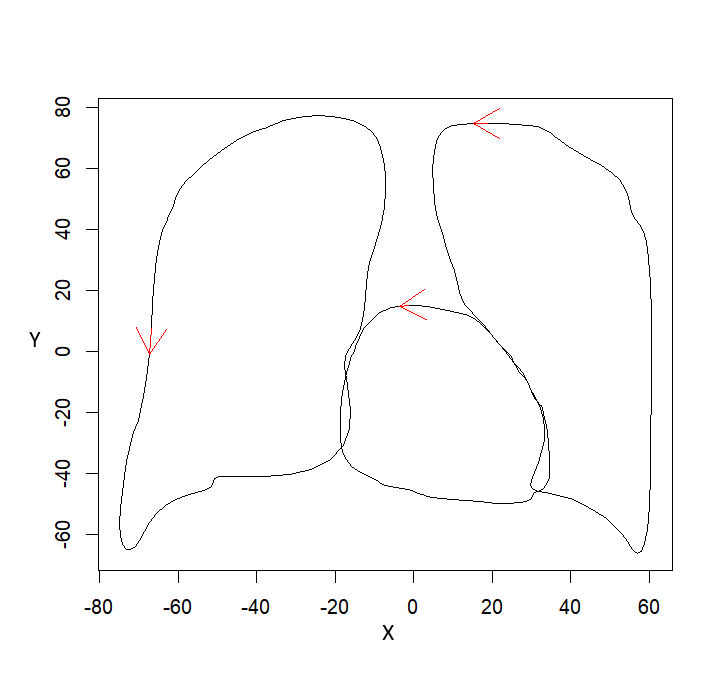}
    \caption{The multivariate ($p=3$) curve $\mathbf{c}^0$ used as the template for generating the synthetic data. Each closed curve represents the contour of an organ, and the red arrow indicates the starting point ($t=0$) used to traverse the contour and define the coordinate functions.}
    \label{ff}
\end{figure}

To this end, we construct a synthetic dataset by deforming a single observation from the CheXmask dataset \citep{gaggion2024chexmask}, which provides multivariate planar curves ($p=3$) corresponding to the contours of the heart and both lungs. We randomly select one observation, smooth it using $M=22$ Fourier basis functions, and center it. This produces a reference curve, denoted $\mathbf{c}^0$, which serves as the template for generating the synthetic data. The curve $\mathbf{c}^0$ can be written as
$$
\mathbf{c}^0=Vec(c_1^0,c_2^0,c_3^0), \quad \text{with} \quad c_j^0 = b_j^0 + a_j^0\boldsymbol{\phi}, \quad j=1,2,3,
$$
where $b_j^0\in\mathbb{R}^{2}$ and $a_j^0\in\mathbb{R}^{2\times 22}$ are coefficient matrices. The resulting function $\mathbf{c}^0$ is illustrated in Figure~\ref{ff}. In particular, the red arrows indicate the starting point ($t=0$) used to traverse each contour.

We then generate pre-shapes $\mathbf{c}_1^*,\ldots,\mathbf{c}_n^*$ by applying random rotation and reparametrization deformations to perturbed versions of $\mathbf{c}^0$. Specifically, we define $\mathbf{c}_{i}^*=Vec(\mathbf{c}_{i1}^*,\mathbf{c}_{i2}^*,\mathbf{c}_{i3}^*)$ with:
\begin{equation}
    \mathbf{c}_{ij}^*= \kappa_i \mathbf{O}_{\theta_i} \left(b_j^0+ {\tilde a}_{ij} \boldsymbol{\phi}\circ \gamma_{\delta_{ij}}\right), \quad j=1,2,3,
    \label{eqxx}
\end{equation}
where $\theta_i \sim \mathcal{U}([0, 2\pi])$, $\delta_{ij} \sim \mathcal{U}([0, 1])$, $\text{Vec}({\tilde a}_{ij}) \sim \mathcal{N}(\text{Vec}(a^0_j), \sigma^2 \mathbf{I}_{44})$, and where $\kappa_i= \left[\sum_{j=1}^3\left(\norm{b^0_j}^2+\norm{{\tilde a}_{ij}}^2\right)\right]^{-1/2}$ ensures normalization. In this construction, the function $\mathbf{c}^0 / \|\mathbf{c}^0\|_{\mathcal{H}^p}$ corresponds to the Fréchet mean of the simulated dataset.

\begin{figure}[h]
    \centering
    \begin{tabular}{c c c c }
         $\sigma=0.1$&\includegraphics[scale=.25, align=c]{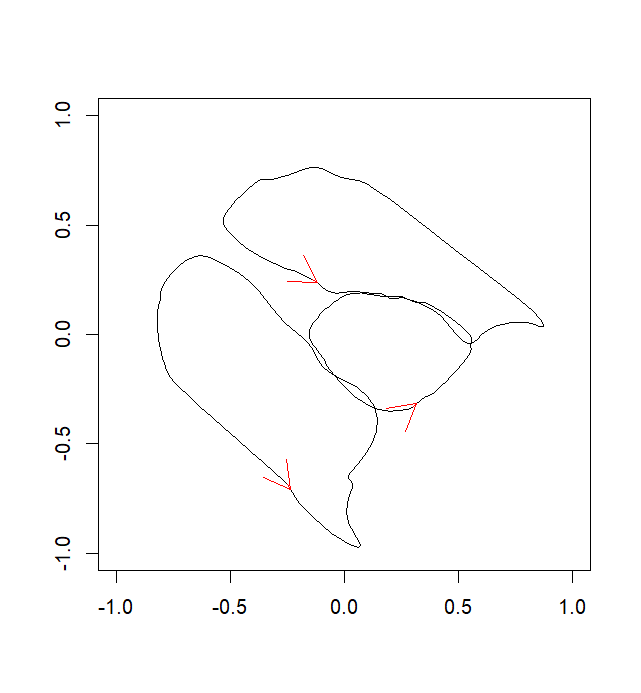}& \includegraphics[scale=.25, align=c]{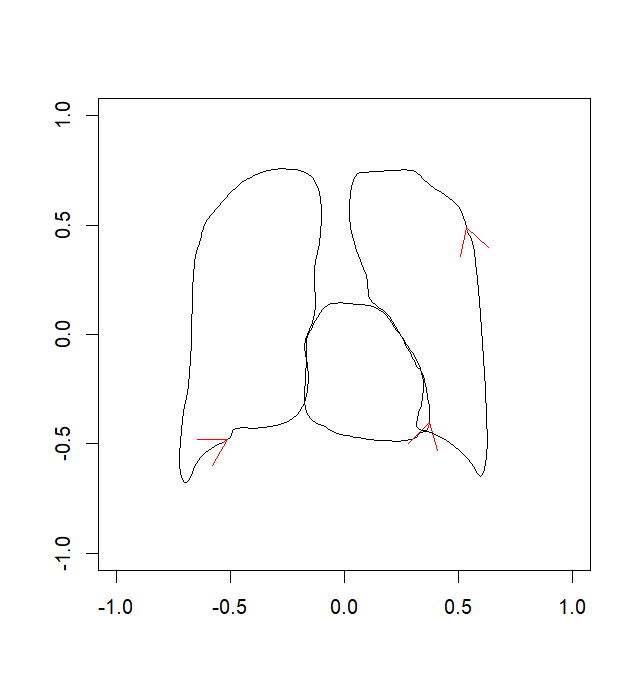} & \includegraphics[scale=.25, align=c]{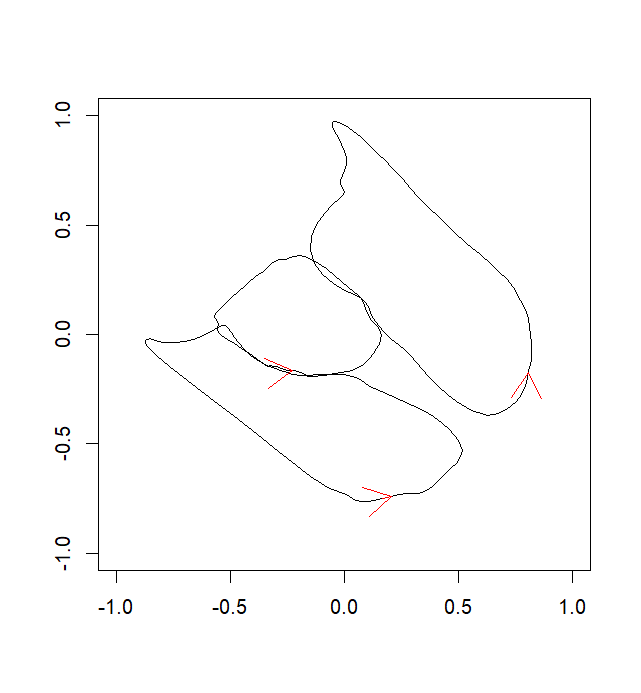} \\
         $\sigma=0.5$& \includegraphics[scale=.25, align=c]{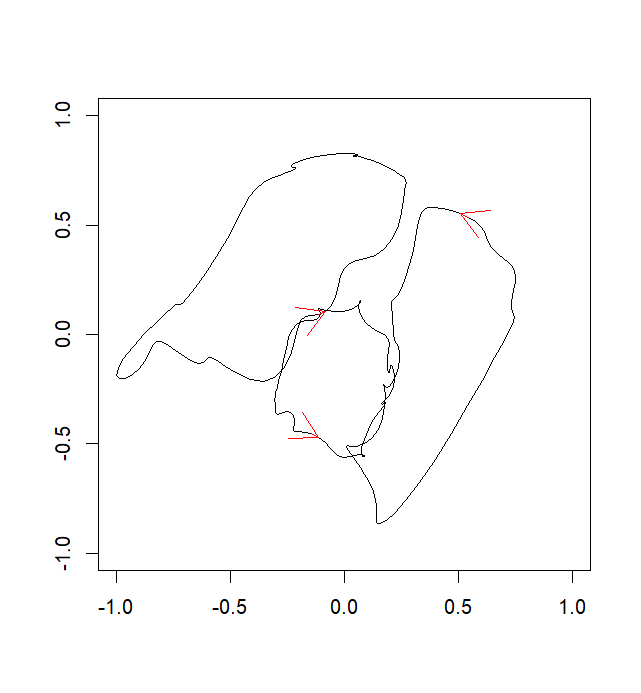}& \includegraphics[scale=.25, align=c]{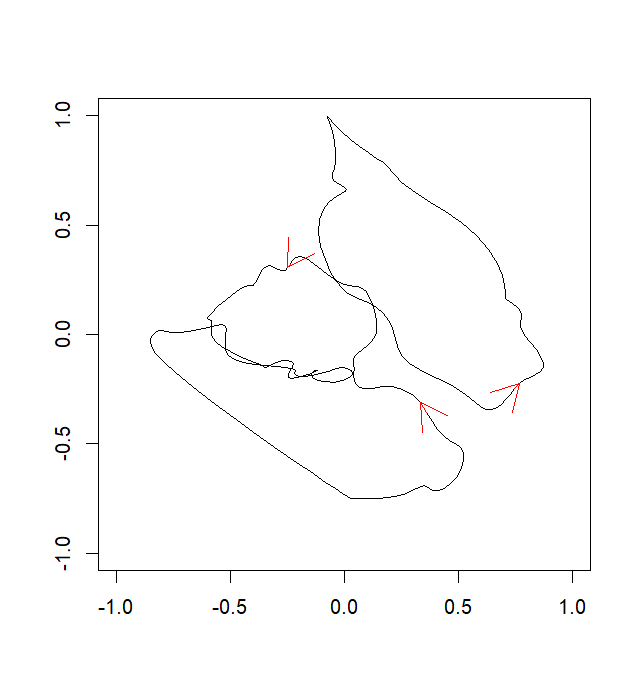} & \includegraphics[scale=.25, align=c]{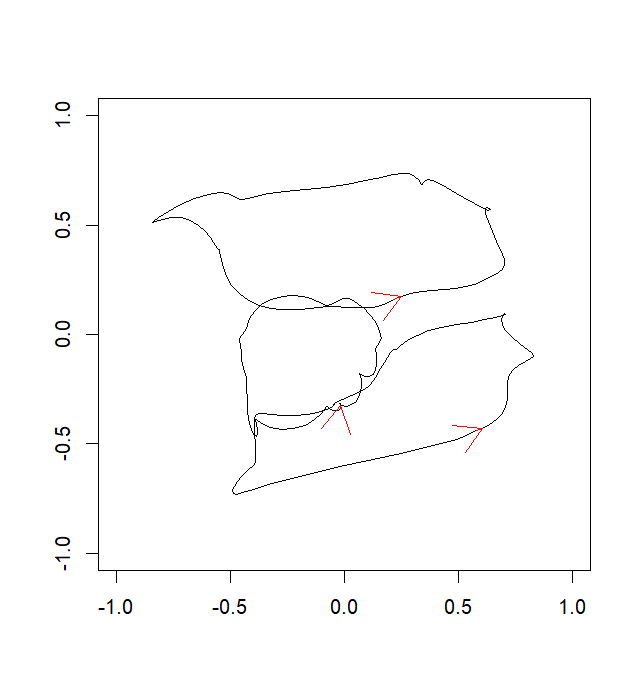} \\
         $\sigma=1$&  \includegraphics[scale=.25, align=c]{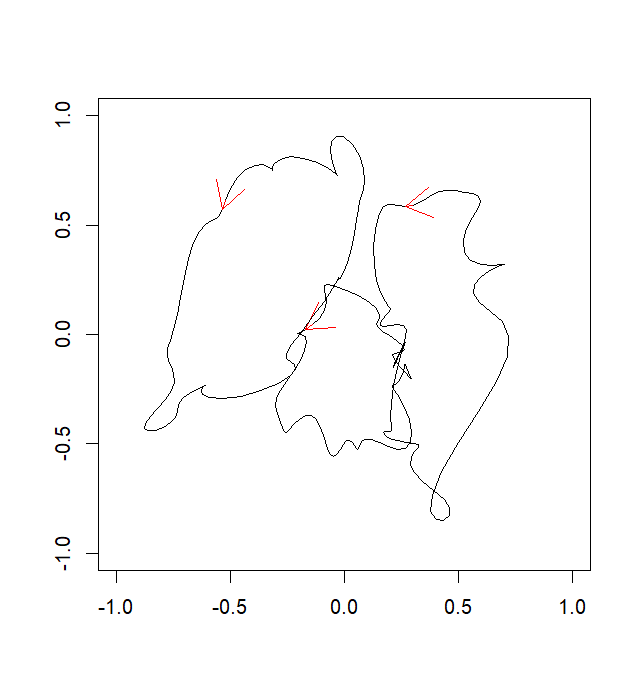}& \includegraphics[scale=.25, align=c]{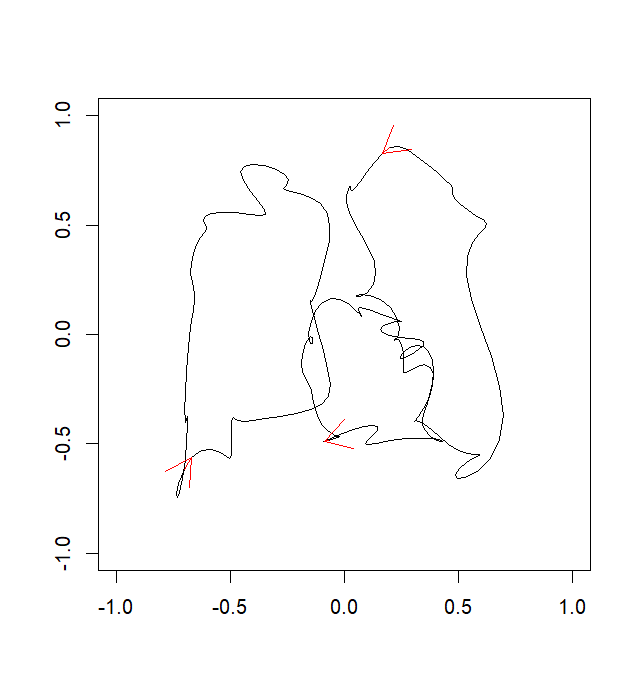} & \includegraphics[scale=.25, align=c]{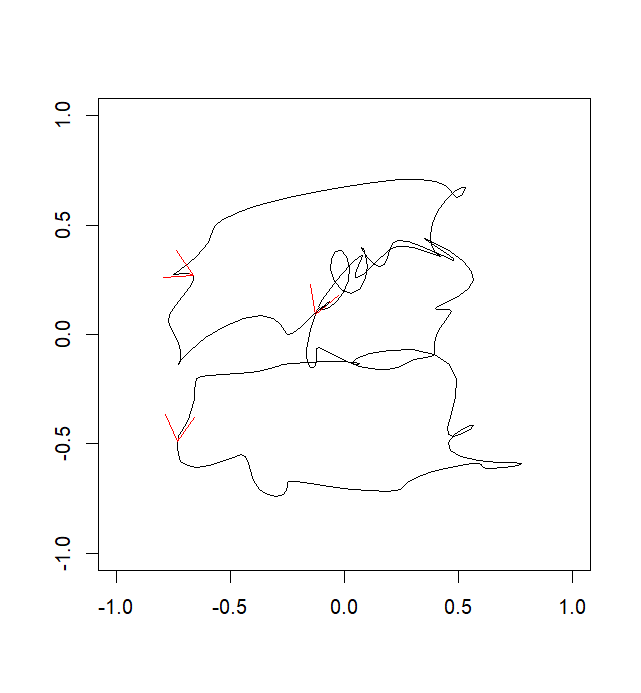}
    \end{tabular}
    \caption{Examples of three pre-shapes generated by applying random rotations and reparametrizations to perturbed versions of the template $\mathbf{c}^0$ under three levels of the perturbation parameter $\sigma$. The rotations change the global orientation of the curves, while the reparametrizations shift the starting points along the contours, illustrated by the red arrows indicating the starting point ($t=0$) used to traverse the curves.}
    \label{align-scen}
\end{figure}

To assess the robustness of the alignment procedures under various
realistic conditions, we generate $n=500$ curves under different levels of the perturbation parameter $\sigma \in \{0.1, 0.5, 1.0\}$. Increasing $\sigma$ produces pre-shapes that deviate progressively from the template $\mathbf{c}^0$, as illustrated in Figure~\ref{align-scen}.

For each of the three datasets, we estimate the deformation parameters $\theta_i$ and $\boldsymbol{\delta}_i = (\delta_{i1}, \delta_{i2}, \delta_{i3})$ associated with the pre-shape $\mathbf{c}_{i}^*$ by applying the multivariate ICF alignment, described in Section~\ref{ficp}, with the template $\bar{\mathbf{C}}$ set to $\boldsymbol{\mu} = \mathbf{c}^0 / \|\mathbf{c}^0\|_{\mathcal{H}^p}$, and we use $M=22$ for the Fourier representation. To improve robustness, we run the ICF algorithm from $5$ random initializations of $\boldsymbol{\delta}_i$ and retain the best one.

To quantitatively assess the accuracy of the estimated deformation parameters, while accounting for the periodic nature of rotations and reparametrizations, we compute the \textit{cyclic mean squared error} (cMSE). For the rotation, it is defined as
\begin{equation*}
    \text{cMSE}_\theta=\frac{1}{500} \sum_{i=1}^{500} \norm{  \begin{pmatrix}
    \cos(\theta_i)\\ 
    \sin(\theta_i)
\end{pmatrix}-\begin{pmatrix}
    \cos(\hat{\theta}_i)\\ 
    \sin(\hat{\theta}_i)
\end{pmatrix} }_2^2,
\end{equation*}
while, separately for each component $j=1,2,3$, the reparametrization cMSE is
\begin{equation*}
\text{cMSE}_{\delta_j} = \frac{1}{500} \sum_{i=1}^{500} \norm{  \begin{pmatrix}
    \cos(2\pi \delta_{ij} )\\ 
    \sin(2\pi \delta_{ij} )
\end{pmatrix}-\begin{pmatrix}
    \cos(2\pi \hat{\delta}_{ij} )\\ 
    \sin(2\pi \hat{\delta}_{ij})
\end{pmatrix} }_2^2.
\end{equation*}

The results are summarized in Table~\ref{res-align}. The low cMSE values across all values of $\sigma$ demonstrate the high accuracy of our alignment procedure, even when the deformation noise is substantial. This supports the reliability of the ICF method for estimating deformation parameters.

\begin{table}[ht]
\centering
\begin{tabular}{lcccc}
  \hline
 $\sigma$& $\text{cMSE}_\theta$ & $\text{cMSE}_{\delta_1}$ & $\text{cMSE}_{\delta_2}$ & $\text{cMSE}_{\delta_3}$  \\ 
  \hline
 0.1& 8.40e-07 & 3.62e-04 & 3.40e-04 & 3.45e-04 \\ 
 0.5& 1.78e-05 & 5.25e-04 & 4.17e-04 & 3.98e-04 \\ 
 1.0 & 6.55e-05 & 1.31e-03 & 6.14e-04 & 5.69e-04 \\  
   \hline
\end{tabular}
\caption{Cyclic mean squared error (cMSE) of the estimated rotation and reparametrization parameters, for different levels of the perturbation parameter $\sigma$.}
\label{res-align}
\end{table}

\begin{figure}[h!]
\centering
\begin{tabular}{l c c c}
  & $
    \mathbf{c}^*_i
$ & $
    \hat{ \tilde{\mathbf{c}} }_i
$ \\  
      & \includegraphics[scale=.2]{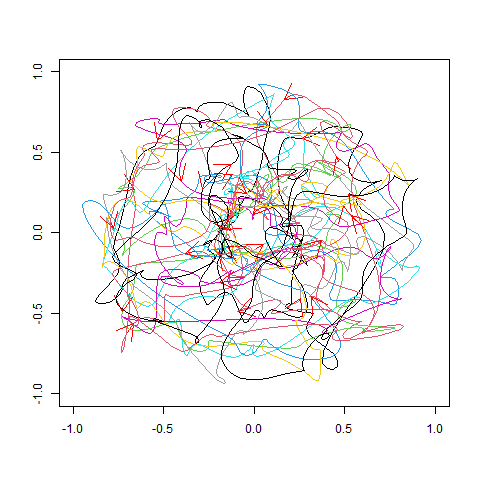}&      \includegraphics[scale=.2]{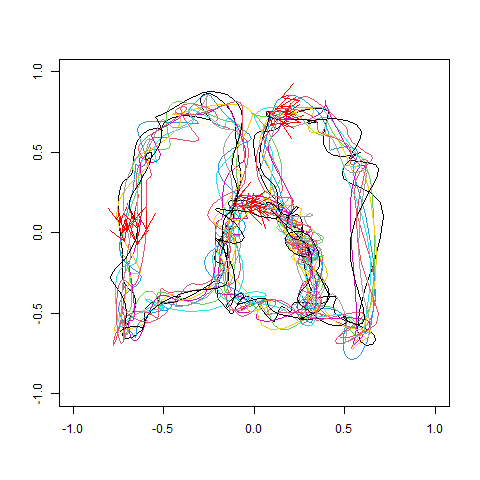}  \\
 (H) & \begin{tabular}{c|c}
 $X$ & $Y$ \\  
            \includegraphics[align=c, scale=.15]{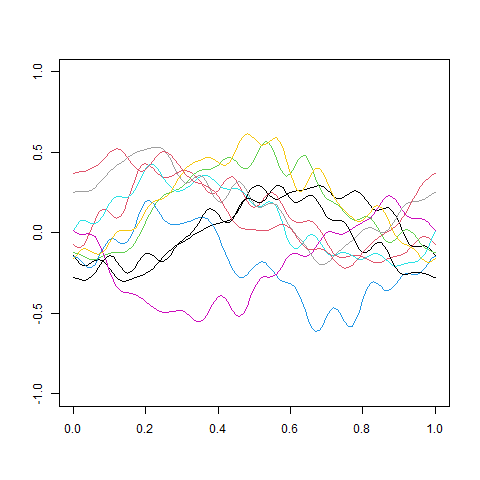}& \includegraphics[align=c, scale=.15]{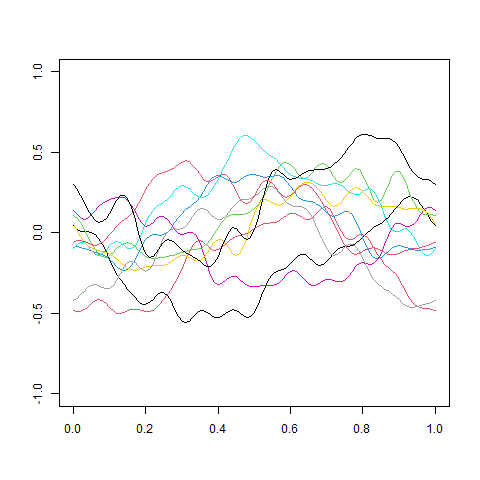}  \\
         \end{tabular} &  \begin{tabular}{c|c}
          $X$ & $Y$ \\ 
              \includegraphics[align=c, scale=.15]{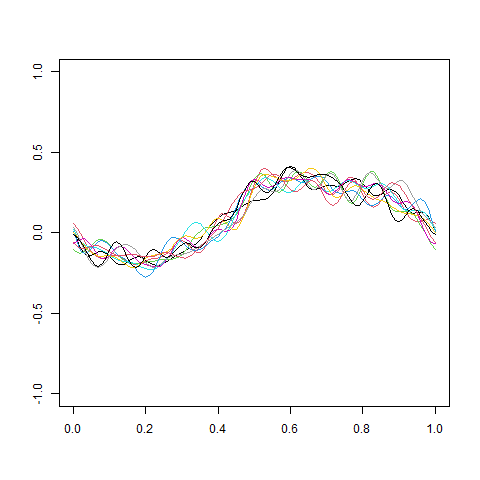}& \includegraphics[align=c, scale=.15]{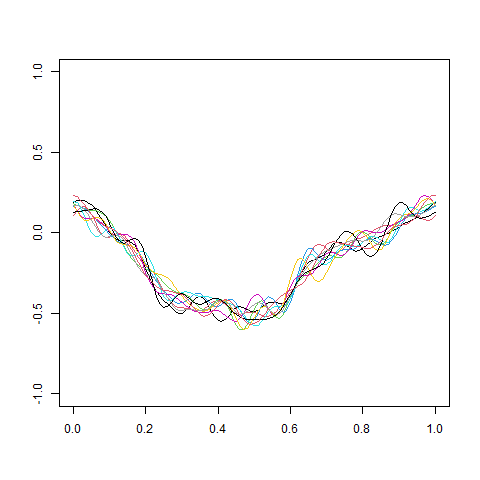}  \\
         \end{tabular} \\ 
     (LL)     &\begin{tabular}{c|c} 
              \includegraphics[align=c, scale=.15]{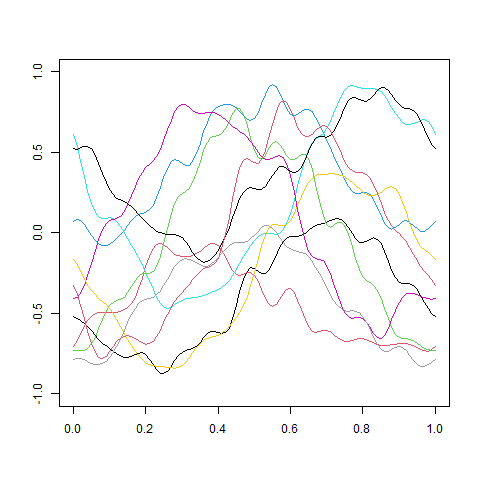}& \includegraphics[align=c, scale=.15]{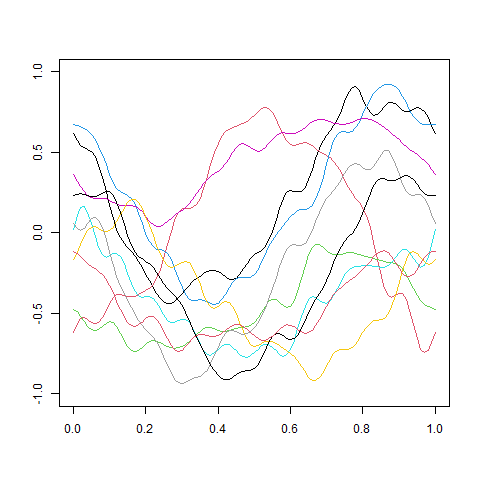}  \\
         \end{tabular} &  \begin{tabular}{c|c}
              \includegraphics[align=c, scale=.15]{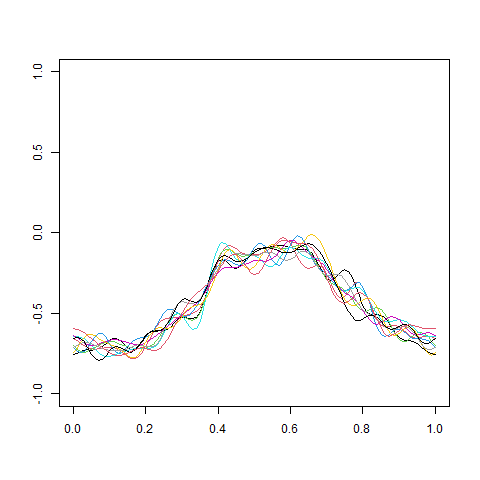}& \includegraphics[align=c, scale=.15]{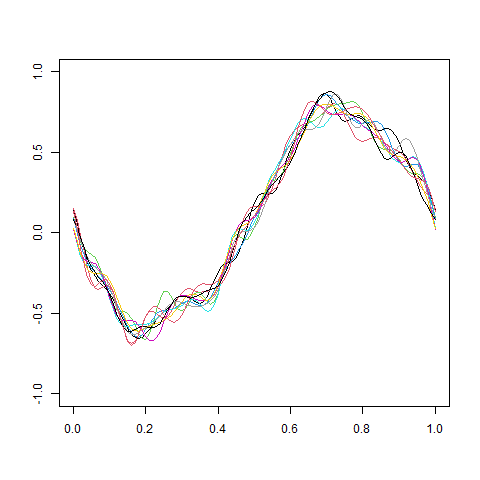}  \\
         \end{tabular}\\ 
          (RL) &\begin{tabular}{c|c}
              \includegraphics[align=c, scale=.15]{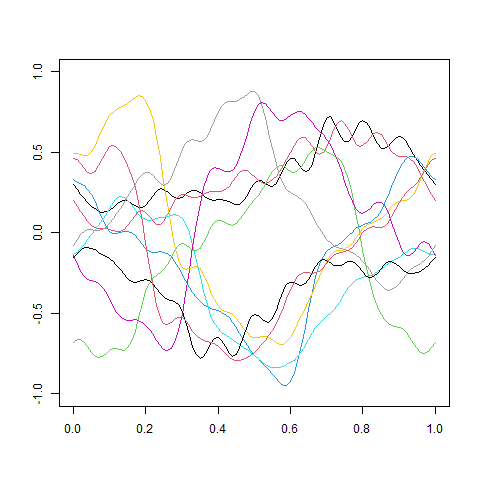}& \includegraphics[align=c, scale=.15]{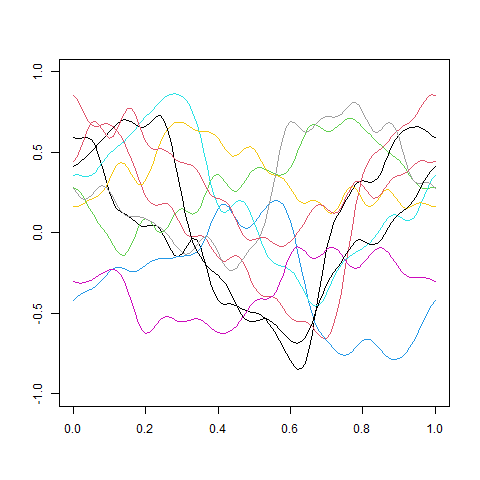}  \\
         \end{tabular} &  \begin{tabular}{c|c}
              \includegraphics[align=c, scale=.15]{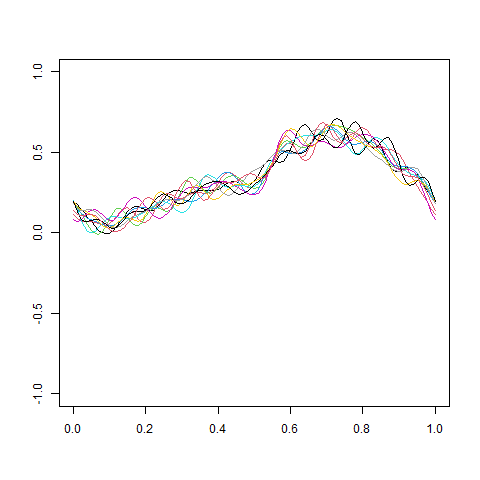}& \includegraphics[align=c, scale=.15]{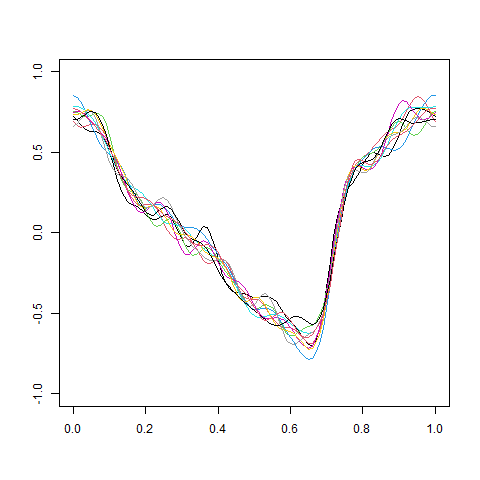}  \\ 
         \end{tabular}
\end{tabular}
\caption{Left panel: a subsample of simulated pre-shapes for $\sigma=1$. Right panel: aligned shapes obtained with the ICF method. The three components of each multivariate curve are the heart (H), the left lung (LL) and the right lung (RL).}
\label{align-scen2}
\end{figure}

Figure~\ref{align-scen2} further illustrates this result by displaying a subsample of simulated pre-shapes for $\sigma=1$ (left panel) along with their corresponding aligned shapes (right panel). A visual inspection confirms the successful recovery of the aligned shapes.

\subsection{Classification of multivariate shapes: a cardiomegaly detection problem}\label{class}

We now illustrate the practical relevance of the proposed methodology by applying it to a real-world classification problem: detecting cardiomegaly from chest X-ray images. Cardiomegaly refers to an abnormal enlargement of the heart, which manifests on radiographic images as the heart occupying more than half of the thoracic width. This condition is clinically important, as it may indicate underlying heart diseases such as heart failure or arrhythmia. 

Our analysis is based on the CheXmask dataset \citep{gaggion2024chexmask}, which contains segmented masks derived from several large-scale chest X-ray databases. We focus on the ChestX-ray8 subset \citep{wang2017}, which provides pathology labels that allow supervised learning.

\begin{figure}[H]
    \centering
    \begin{tabular}{c c c c}
     (a) & \includegraphics[scale=.20, align=c]{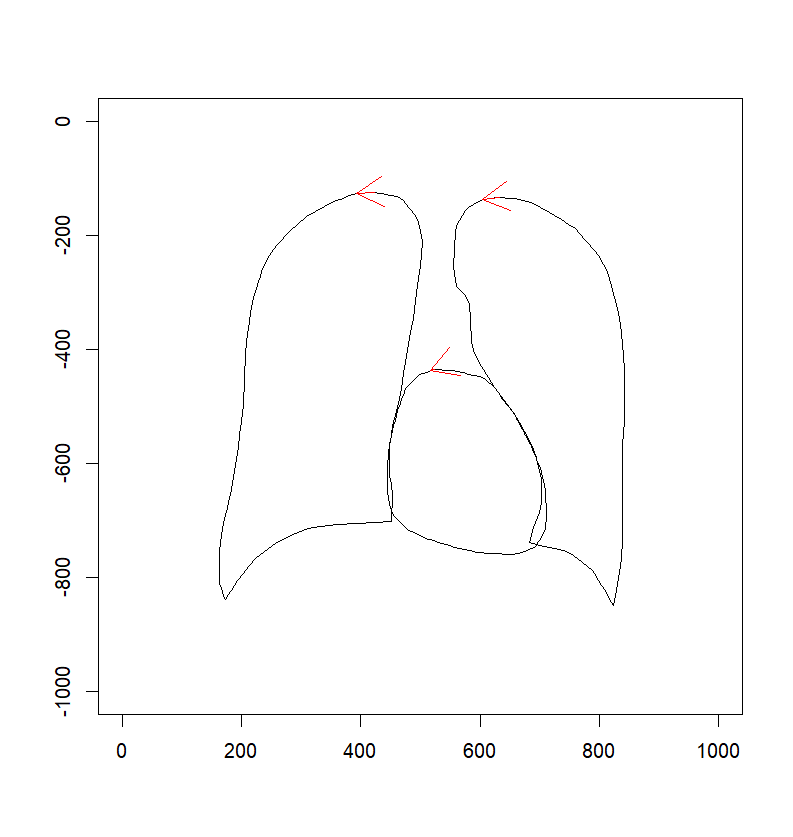}    &  \includegraphics[scale=.20, align=c]{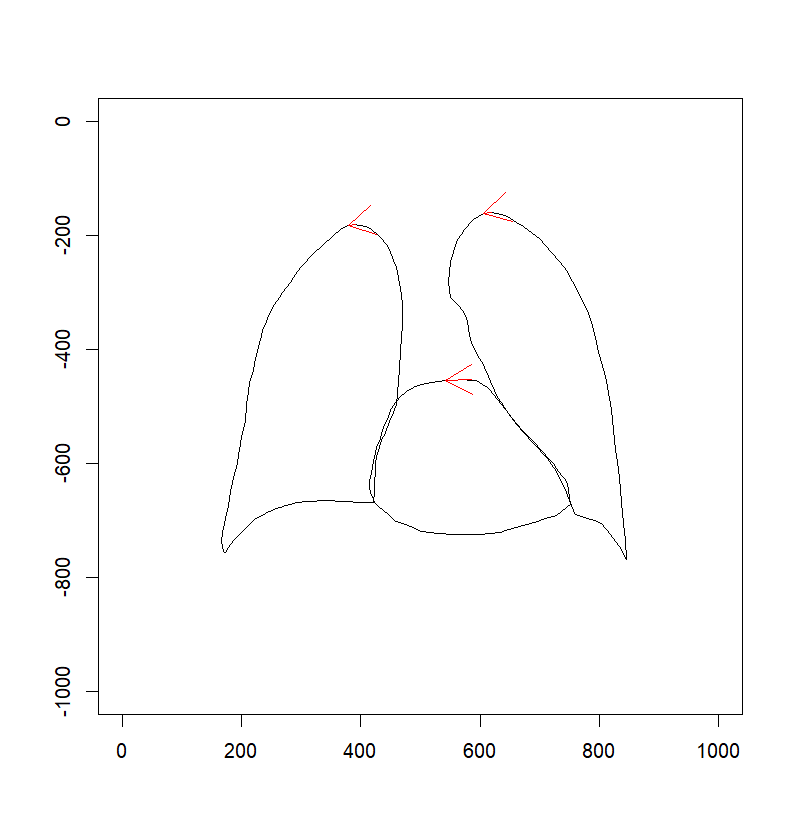} & \includegraphics[scale=.20, align=c]{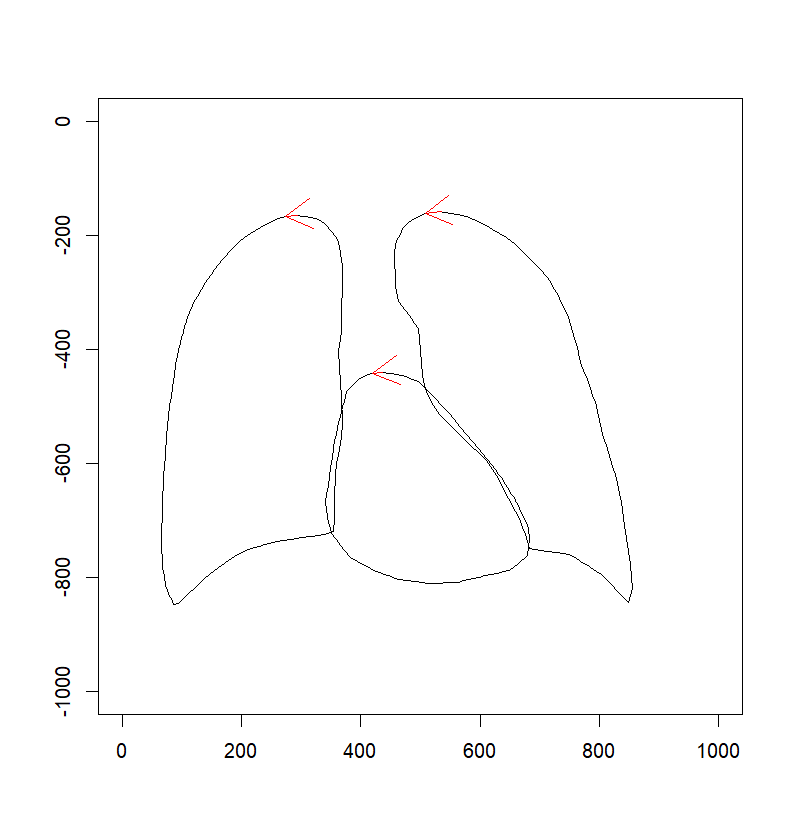}   \\
        (b) & \includegraphics[scale=.2, align=c]{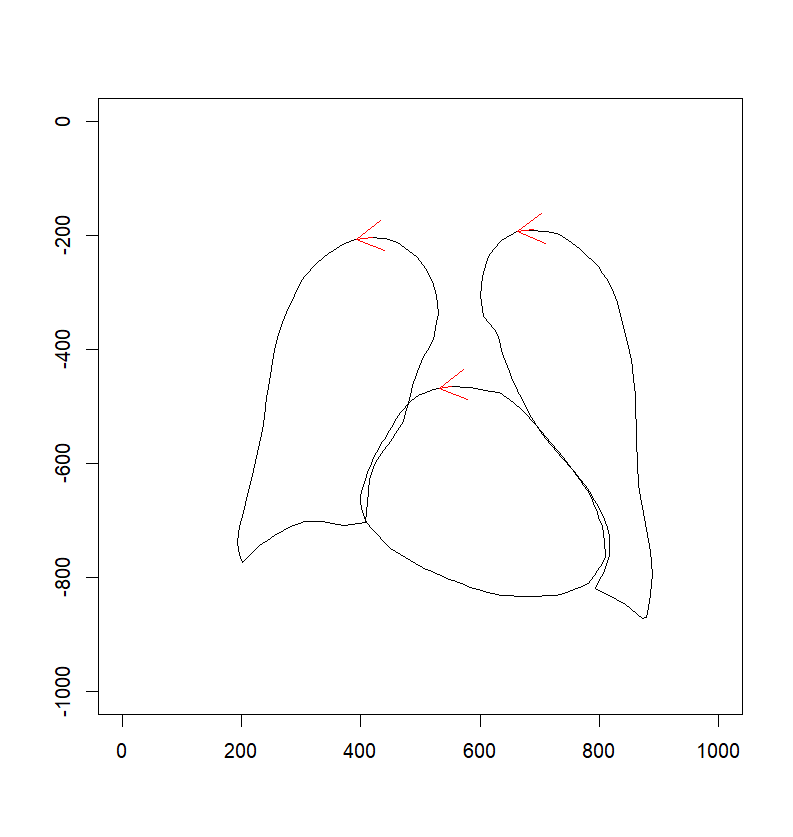}    &  \includegraphics[scale=.2, align=c]{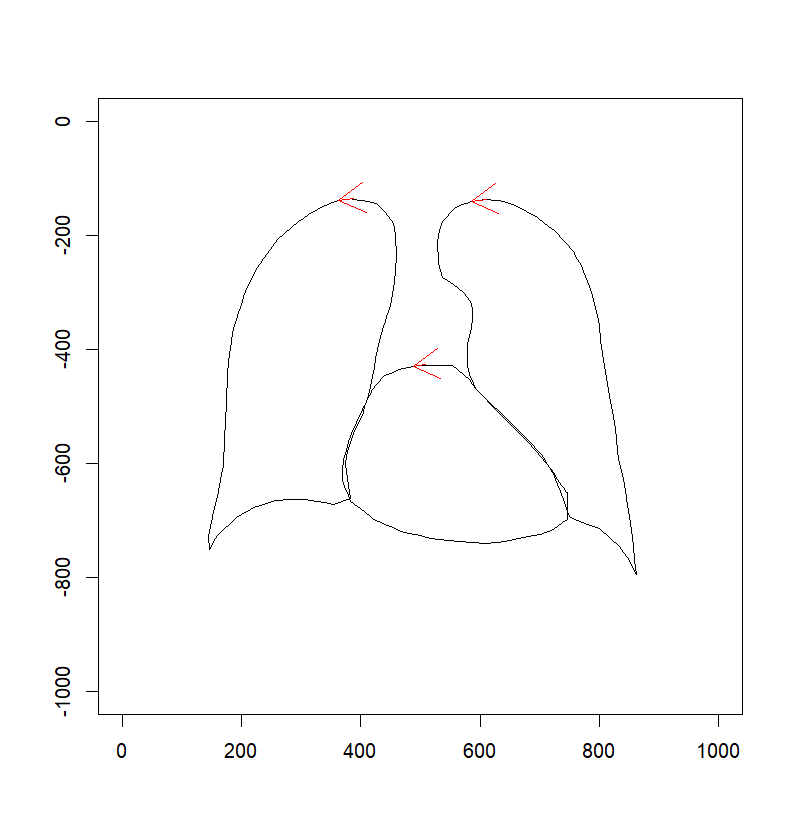} & \includegraphics[scale=.2, align=c]{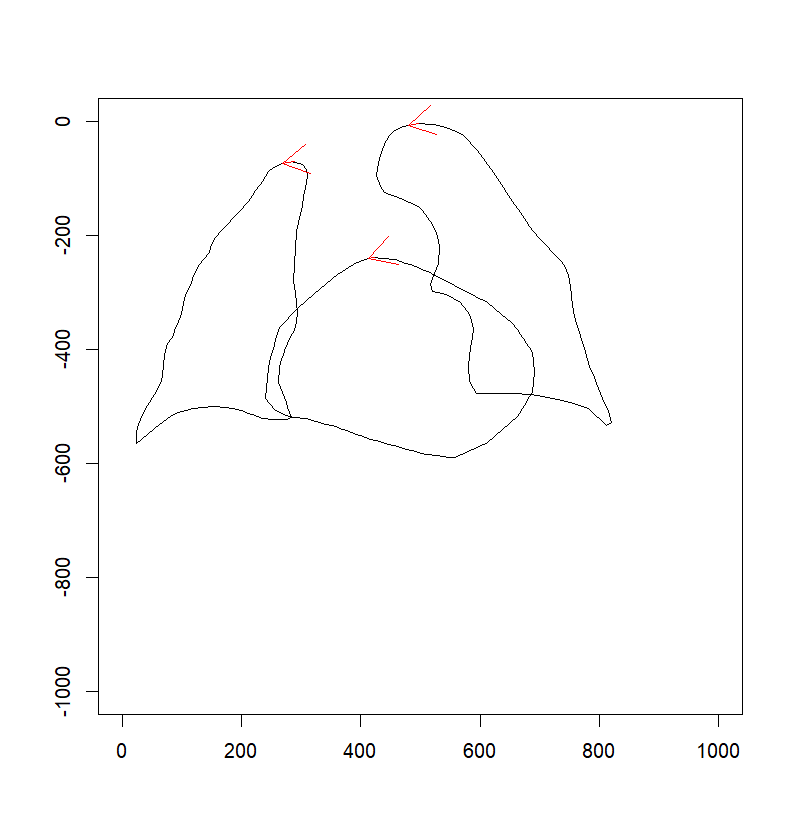}  
    \end{tabular}
       \caption{Examples of multivariate planar curves extracted from chest X-ray images: (a) healthy patients and (b) patients diagnosed with cardiomegaly. Each panel shows the contours of the heart and both lungs for three representative observations. The red arrows indicate the starting point $t=0$.}
    \label{card}
\end{figure}

Each segmented image is represented as a multivariate planar curve with $p=3$ components corresponding to the contours of the heart and the left and right lungs. This representation naturally fits the framework developed in this paper. Figure~\ref{card} illustrates examples of contours extracted from chest X-ray images of healthy patients and patients diagnosed with cardiomegaly. Although cardiomegaly is characterized by an enlargement of the heart relative to the lungs, the visual distinction between the two classes can be subtle. As shown in the figure, there is substantial variability in both the shapes of the organs and their relative deformations across individuals, which makes automated classification a challenging task.

\begin{figure}[H]
    \centering 
     \includegraphics[scale=.4, align=c]{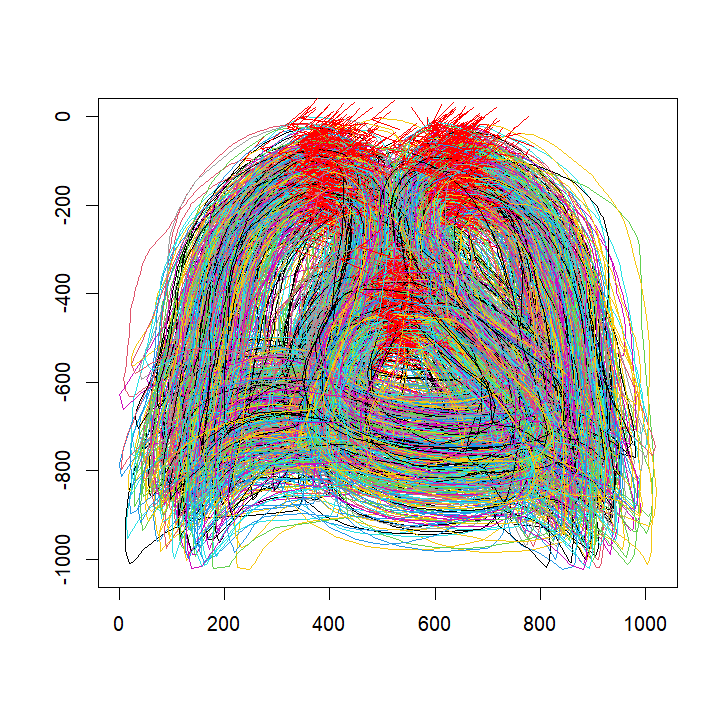}
\caption{The considered subsample of $300$ multivariate planar curves $\mathbf{c}_i$ from the chest X-ray dataset \citep{gaggion2024chexmask}. Different observations are identified using different colors and the red arrows represent the starting
point ($t=0$) for each coordinate function. }
\label{ap-pres}
\end{figure}

From the original dataset, we extract a balanced subset of $n=300$ images, consisting of the $150$ X-rays labeled as ``cardiomegaly'' and $150$ randomly selected healthy X-rays labeled as ``no finding''. The resulting dataset, denoted $ \mathbf{c}_1,\ldots, \mathbf{c}_n$, is illustrated in Figure \ref{ap-pres} and can be found in our online code repository (\url{https://github.com/imoindjie/Shape-MFDA}).

It is important to note that the curves provided in \cite{gaggion2024chexmask} have already been preprocessed to reduce translation, rotation, and scale variability using standard image registration techniques. However, visual inspection of Figure \ref{ap-pres} reveals that residual misalignments remain, particularly in the location of the starting points used to traverse the contours.

To study the robustness of our classification pipeline, we consider two experimental scenarios:

\paragraph{Scenario 1 - Pre-aligned data} We directly use the pre-aligned curves $\mathbf{c}_i,  i=1,\ldots,n$, as provided in the dataset.
\paragraph{Scenario 2 - Misaligned data} To assess the benefit of alignment in a controlled setting, we introduce artificial deformations to each observation. We set $\mathbf{c}_i$, $i=1,\ldots,n$, as
\begin{equation}
    \mathbf{c}_i \leftarrow (\mathbf{I}_3\otimes\mathbf{O}_{2\pi \zeta_i})\mathbf{ c}_i \boldsymbol{\circ} \boldsymbol{\gamma}_{\boldsymbol{\delta}_i}, \textrm{ with } \boldsymbol{\delta}_i=(\delta_{i1},\delta_{i2},\delta_{i3})\in [0, 1]^3,
\end{equation}
where $\zeta_i \sim \mathcal{U}([0,1])$ and each reparametrization component $\delta_{ij} \sim \mathcal{U}([0,1]),$ for $j=1,2,3$. This setting allows us to evaluate how well the alignment methods handle misaligned input data.

\subsubsection{Classification models}

As introduced in Section~\ref{sec_classi}, the classification framework uses a predictor of the form $\mathcal{L}_{\boldsymbol{\mu}}(\tilde{\mathbf{C}})$, which, we recall, corresponds to the projection of the multivariate shape $\tilde{\mathbf{C}}$ on the tangent space $\mathcal{T}_{{\boldsymbol{\mu}}}$ where $\boldsymbol{\mu}$ is the Fréchet mean of $\tilde{\mathbf{C}}$. 

To assess the benefits of our proposed multivariate approach, we compare three ways of defining the predictors:
\begin{itemize}
\item \textit{OUR} - our proposed approach: we use $\mathcal{L}_{\hat{\boldsymbol{\mu}}}(\hat{\tilde{\mathbf{c}}}_i)$, $i=1,\ldots,n$, as predictors, where the shapes and Fréchet mean are estimated using the algorithm described in Section \ref{sec_iter_algo}. This method jointly models the three anatomical components.
\item \textit{UNI} - univariate approach: each anatomical component is analyzed separately, as in equation \eqref{eq_unie}, and the shapes are estimated following the univariate procedure of \cite{axe1}. We use $(\mathcal{L}_{\hat{{\mu}}_1}(\hat{\mathcal{C}}_{i1}),\mathcal{L}_{\hat{{\mu}}_2}(\hat{\mathcal{C}}_{i2}),\mathcal{L}_{\hat{{\mu}}_3}(\hat{\mathcal{C}}_{i3}))^\top$, $i=1,\ldots,n$, as predictors, where $\mathcal{L}$ is defined for $p=1$ and $\hat{{\mu}}_j$ is the estimated empirical Fréchet mean of $\mathcal{C}_{1j},\ldots,\mathcal{C}_{nj}$, for $j=1,2,3$. This approach ignores the global multivariate structure and treats each organ independently.
\item \textit{CLA} - classical naïve approach: we use the raw observed curves $\mathbf{c}_i$ as predictors, without any alignment. 
\end{itemize}

To ensure that performance differences are not tied to a specific classifier, we evaluate these three predictor sets with four different functional linear classification methods, corresponding to different choices of the link function $\psi$ in \eqref{model_classi}.

Two methods are based on penalized group logistic regression. Following the group lasso framework \citep{meier2008group} (see also \cite{godwin2013} for its functional extension), functional predictors are partitioned into predefined groups before applying the penalty:

\begin{itemize}
\item[{GL1}:] Three groups, each corresponding to one anatomical component $c_{j}=(x_{j},y_{j})^\top, j=1,2,3$. The penalty can identify whether the heart or one of the lungs has no discriminative power in the model.
\item[{GL2}:] Six groups, each corresponding to a single coordinate function $x_{j}$ or $y_{j}$, for $j=1,2,3$. This allows for finer selection, identifying specific coordinates that are non-informative.
\end{itemize} 

The penalty parameter $\lambda$ is selected, independently for the two methods, as the value that minimizes the classification error over the set $ \left \{0.96^{l}\lambda_{\max}, \ l = 0,1, \ldots, 148\right\} \cup \left\{0\right\}$,
where $\lambda_{\max}$ is the smallest value for which the penalty term vanishes. 

The two other considered methods rely on linear discriminant analysis where the discriminant coefficient function is estimated via partial least squares (PLS) and principal component regression (PCR) (\cite{upls}, \cite{mfpls}). For both, the number of components is chosen from the set $\{1,2, \ldots, 6\times 22-1\}$ to minimize classification error.

\subsubsection{Results}

We evaluate the performance of each classification strategy using a 10-fold cross-validation procedure. The results for both scenarios are reported in Table~\ref{table_res}.

In Scenario 1 (aligned curves), the results show that although the four classification procedures (GL1, GL2, PLS, PCR) lead to varying accuracy levels, the \textit{OUR} and \textit{CLA} approaches achieve almost identical performance across all methods. This confirms that, when the multivariate curves are already aligned, the advantage of a shape-based preprocessing step is minimal. The \textit{UNI} approach performs slightly worse, likely because it treats the heart and each lung as independent univariate curves, so that their relative spatial relationships are ignored. Nevertheless, its accuracy remains relatively high, underlining the strong discriminative power of shape information alone.

In Scenario 2 (misaligned curves), the situation changes markedly. The classical \textit{CLA} accuracies collapse to $53$--$57\%$, close to random guessing, which highlights the detrimental impact of ignoring alignment in multivariate shape classification. The two shape-based methods, by contrast, remain as accurate as in Scenario 1, confirming their robustness to rotation and reparametrization. Among them, our proposed multivariate approach \textit{OUR} is the best for every classifier and outperforms the univariate \textit{UNI} by $3$ to $11$ accuracy points. This gap reflects the core benefit of our framework: modeling the three contours jointly preserves the inter-component information, relative position, scale and orientation, that a separate, univariate treatment discards, yet that is discriminative for cardiomegaly.

Overall, these results emphasize two main points: (i) when data are aligned, our approach can match the performance of classical methods, and (ii) in settings with misaligned data, shape-based methods, particularly our proposed multivariate framework, are more reliable.

\begin{table}[H]
    \centering
    \begin{tabular}{llcccc}
        \hline
        Scenario & Method & GL1 & GL2 & PLS & PCR \\
        \hline
        \multirow{3}{*}{Scenario 1} & \textit{OUR} & 79.65 & 77.34 & 85.22 & {82.94} \\
                                    & \textit{UNI} & 71.73 & 71.94 & 78.83 & 73.36 \\
                                    & \textit{CLA} & {80.00} &{78.27} & {85.25} & 80.60 \\
        \hline
        \multirow{3}{*}{Scenario 2} & \textit{OUR} &{80.50} & {78.39} & {85.99} & {82.77} \\
                                    & \textit{UNI} & 77.37 & 74.11 & 77.92 & 71.67 \\
                                    & \textit{CLA} & 57.11 & 54.94 & 55.15 & 53.13 \\
        \hline
    \end{tabular}
    \caption{Classification accuracies (\%) obtained by 10-fold cross-validation, for the three predictor sets (\textit{OUR}, \textit{UNI}, \textit{CLA}) and the four classifiers (GL1, GL2, PLS, PCR), under the two scenarios.}
    \label{table_res}
\end{table}

 \section{Conclusion}\label{sec_conclusion}

This work extends statistical shape analysis to multiple objects by adopting a functional data analysis perspective. As in the univariate case, the observed multivariate planar curve is viewed as a deformed version of an unobserved latent variable of interest: its shape. The novelty of our work lies in treating the multiple contours as a single multivariate planar curve, which allows several objects to be studied jointly while preserving the geometric information they share. After defining a suitable and intuitive model in this setting, we introduce estimation procedures for the deformation and shape variables. In particular, we extend the Iterative Closest Function (ICF) procedure of \cite{axe1} to jointly estimate the rotation and reparametrization effects of the multiple contours, from which the shapes are recovered through inverse transformations. We then describe how to summarize and analyze these shapes, through their Fréchet mean and a projection onto the associated tangent space, so that they can serve as predictors in functional classification models. \par  
The numerical experiments demonstrate the proficiency of our method in estimating the deformation parameters and recovering the shapes across diverse scenarios. In the cardiomegaly detection problem, they further show that classifying multivariate shapes improves accuracy over both a univariate shape analysis, which treats each contour separately, and a classical approach that feeds the raw, unaligned curves to the same classifiers, particularly in the setting where images are not pre-aligned. This advantage stems from modeling the contours jointly, which preserves the inter-component information that is discriminative for the task. The proposed representation for multivariate planar curves is moreover parsimonious, and we believe that the low computational cost and transparency of the proposed shape-based methods, compared to pixel-based approaches, make them appealing alternatives for image analysis when objects are well identified. \par 
Indeed, the availability of segmented images makes our approach readily applicable; however, the quality of current image segmentation models remains a limitation for analyzing shapes in more complex images. As image segmentation and contour detection constitute active research domains, future work should focus on the challenge of extracting multivariate planar curves directly from colored images. For this aim, we could rely on the rich literature within the computer vision community (see e.g., \cite{kass1988}, \cite{ibrahim2020}).  \par 
Moreover, the underlying idea of this work, which defines an image solely through the objects' contours, can be limiting in many application cases. Indeed, in addition to shapes, colors in images are strongly informative in the classification task (see e.g., \citep{gowda2018colornet}, \cite{funt2018does}). Hence, investigating a fully functional approach to account for these variations is of interest. An intuitive idea is to shift the perspective from shapes to surfaces, where pixels represent a discrete sample over such surfaces. Future work should focus on keeping the procedure interpretable and the image representation as parsimonious as possible, to avoid requiring more computing resources than pixel-based methods. The resulting methods could then be used to fairly assess the advantages of using pixels versus shapes and colors as inputs in prediction problems.
 \par 
\newpage 

\bibliography{refs}
\bibliographystyle{apalike}

\end{document}